\documentclass{article}
\usepackage[utf8]{inputenc}
\usepackage{graphicx}
\usepackage[]{algpseudocode}
\usepackage{anysize}
\usepackage{todonotes}
\usepackage{mathtools}
\usepackage{csquotes}
\usepackage{hyperref}
\usepackage{caption,subcaption}
\usepackage{listings}
\usepackage{amsfonts}
\usepackage{xcolor,colortbl}
\usepackage{tabularx}

\definecolor{bid_green}{rgb}{0.6235,0.835,0.619}
\definecolor{ask_red}{rgb}{0.898,0.592,0.592}
\colorlet{punct}{red!60!black}
\definecolor{background}{HTML}{EEEEEE}
\definecolor{delim}{RGB}{20,105,176}
\colorlet{numb}{magenta!60!black}

\newcommand{\ModelName}{XChange}

\newcommand{\MsgOrder}{\texttt{Order}}
\newcommand{\MsgCancelOrder}{\texttt{CancelOrder}}
\newcommand{\MsgMatch}{\texttt{Match}}
\newcommand{\MsgRejectMatch}{\texttt{RejectMatch}}
\newcommand{\MsgTrdProp}{\texttt{TradeProposal}}
\newcommand{\MsgTrdNegotiate}{\texttt{Negotiate}}
\newcommand{\MsgTrdAccept}{\texttt{TradeAccept}}
\newcommand{\MsgTrdReject}{\texttt{TradeReject}}
\newcommand{\MsgPartialAgreement}{\texttt{PartialAgreement}}
\newcommand{\MsgAgreement}{\texttt{Agreement}}
\newcommand{\MsgPayment}{\texttt{Payment}}
\newcommand{\MsgPartialTradeDone}{\texttt{PartialTradeDone}}
\newcommand{\MsgTradeDone}{\texttt{TradeDone}}

\newcommand{\TROffer}{\textsc{Offer}}
\newcommand{\TRRequest}{\textsc{Request}}

\newcommand{\TRAgreement}{\textsc{Agreement}}
\newcommand{\TRPayment}{\textsc{Payment}}
\newcommand{\TRTradeDone}{\textsc{TradeDone}}

\author{
	Martijn de Vos\\
	\texttt{m.a.devos-1@tudelft.nl}
	\and
	Can Umut Ileri\\
	\texttt{c.u.ileri@tudelft.nl}
	\and
	Johan Pouwelse\\
	\texttt{j.a.pouwelse@tudelft.nl}
}

\title{XChange: A Blockchain-based Mechanism for Generic Asset Trading In Resource-constrained Environments\\ \large Technical Report}

\begin{document}
	
\maketitle

\begin{abstract}
	An increasing number of industries rely on Internet-of-Things devices to track physical resources.
	Blockchain technology provides primitives to represent these resources as digital assets on a secure distributed ledger.
	Due to the proliferation of blockchain-based assets, there is an increasing need for a generic mechanism to trade assets between isolated platforms.
	To date, there is no such mechanism without reliance on a trusted third party.
	
	In this work, we address this shortcoming and present XChange.
	XChange mediates trade of \emph{any} digital asset between isolated blockchain platforms while limiting the fraud conducted by adversarial parties.
	We first describe a generic, five-phase trading protocol that establishes and executes trade between individuals.
	This protocol accounts full trade specifications on a separate blockchain.
	We then devise a lightweight system architecture, composed of all required components for a generic asset marketplace.
	
	We implement XChange and conduct real-world experimentation.
	We leverage an existing, lightweight blockchain, TrustChain, to account all orders and full trade specifications.
	By deploying XChange on multiple low-resource devices, we show that a full trade completes within half a second.
	To quantify the scalability of our mechanism, we conduct further experiments on our compute cluster.
	We conclude that the throughput of XChange, in terms of trades per second, scales linearly with the system load.
	Furthermore, we find that XChange exhibits superior throughput and order fulfil latency compared to related decentralized exchanges, BitShares and Waves.
\end{abstract}

\section{Introduction}
Advancements in wireless technology, sensor networks, and hardware capabilities have resulted in exponential growth of low-resource devices operating within the Internet-of-Things (IoT)~\cite{makhdoom2018blockchain}.
IoT technology holds the potential to address societal challenges such as food security, sustainable agriculture, and renewable transportation~\cite{atzori2017understanding}.
Gartner estimates that by 2020, the Internet-of-Things consists of 20 billion Internet-connected devices worldwide~\cite{hung2017leading}.

Many industries have deployed devices for the global traceability of assets, resources, or services.
For example, a smart meter installed at the home of a consumer keeps track of the energy usage of a specific household and reports these values to the energy provider.
Likewise, a GPS tracker installed in taxis calculates the traveled distance, so the driver knows how much to charge its passengers for a ride.
Such use-cases require an accountability mechanism to securely record the events that influence the state of an asset, e.g., physical movement or ownership change.

One of the solutions that can bring accountability features to the Internet-of-Things is blockchain technology.
Blockchain is a paradigm to create, manage and transfer digital assets in a tamper-proof manner.
This management is achieved by maintaining a distributed ledger that enables secure recording of transactions between digital entities, even within the context of mutual distrust.
The distributed ledger, also called a blockchain, is operated by network participants themselves without any dependency on trusted third parties.
From a business perspective, it fosters collaboration between companies since it provides the means to securely record interactions and share data. 


Various industries are using a blockchain to keep track of assets within their domain.
Representing assets on a distributed blockchain is also called \emph{tokenization}~\cite{TokenizationOP}.
The introduction of the Bitcoin cryptocurrency and the popularity of blockchain technology in general have resulted in a proliferation of different types of digital assets, fragmented across many blockchain implementations.
Currently, almost 200.000 different assets are being managed on the Ethereum blockchain only.\footnote{https://etherscan.io/tokens} 
While there is an abundance of different types of digital assets, there is no generic mechanism to exchange (trade) these assets between isolated blockchain platforms without the involvement of a trusted third party.
We argue that such a mechanism is a growing necessity, as more industries are relying on blockchain technology and tokenization of their resources.


We present \emph{\ModelName{}}, a blockchain-based mechanism for generic asset trading in resource-constrained environments.
In stark contrast to existing approaches, \ModelName{} enables the exchange of digital assets between practically \emph{any} blockchain platform without need for trusted third parties that mediate between market participants.
\ModelName{} is highly applicable in situations where many different assets have to be managed and exchanged.
These situations include peer-to-peer energy trading, supply chain management, and service exchange within the sharing economy.

The key idea of \ModelName{} is to account full trade specifications on a blockchain.
This accounting enables traders to accurately track the progress of ongoing trades.
A trade in \ModelName{} is modelled as a sequence of unilateral payments (asset transfers) between two trading parties.
\ModelName{} prevents a trader from initiating a trade with someone that currently have an obligation to initiate a payment to another trader during an ongoing trade.
This addresses the situation where an adversarial party steals assets by refusing to initiate a payment back to some counterparty after having received some assets.
As a result, an adversary can only commit this kind of fraud once.
To further reduce the gains of adversarial parties, \ModelName{} allows that each trade is divided in multiple payments.
We call this mechanism \emph{incremental settlement}.


The main contribution of this work is five-fold:
\begin{enumerate}
	\item The \ModelName{} \emph{trading protocol} which specifies how cross-chain trade between two parties proceeds, and how fraud conducted by adversaries is limited. (Section \ref{sec:protocol}). 
	\item A lightweight \emph{system architecture} for generic, cross-chain trade (Section \ref{sec:architecture}).
	\item An improvement of TrustChain \cite{otte2017trustchain}, the consensus-less blockchain solution used by \ModelName{}, that enables concurrent transactions and increases scalability (Section~\ref{sec:blockchain_accounting}).
	\item A fully functional, open source \emph{implementation} of the \ModelName{} trading protocol and system architecture (Section \ref{sec:implementation}).
	\item \emph{Experimentation} around the resource usage and scalability of \ModelName{}, conducted on multiple low-resource devices and our compute cluster (Section~\ref{sec:exp_trading_low_devices}, \ref{subsec:scalability_experiment} and \ref{sec:experiment_comparison}).
\end{enumerate}

\section{Background and Problem Description}
This work focuses on how to trade digital assets that are stored on different blockchains, without the need for a trusted market operator that mediates in the trading process.
We first provide background on electronic marketplaces and elaborate how blockchain technology enables decentralized asset marketplaces.
We then formulate the problem description and pose the cardinal research question that this work answers.

\subsection{Electronic Marketplaces}
For decades, the ability to facilitate trading at a global scale has been at the core of many established companies operating on the Internet.
In 1995, Craigslist already offered an unmoderated mailing list where strangers could negotiate, meet, and trade physical goods.
eBay formalized the concept of online trading by introducing a reputation system where buyers and sellers rate each other.
More recently, leading companies acting within the sharing economy, deployed global marketplaces for ride-hailing (Uber), accommodation (AirBnb), freelance labour (TaskRabbit), and many other services.

The primary responsibility of electronic marketplaces is two-fold.
First, market operators have to ensure a quick and effective mediation between supply and demand.
In most electronic marketplaces, participants create supply and demand by submitting \emph{orders} to the market operator.
In general, literature distinguishes between two types of orders: \emph{offers} that indicate the willingness of traders to sell specific assets, and \emph{requests}, created by a party interested in specific assets.
Electronic marketplaces aggregate submitted offers and requests, and match them, based on their specifications.
This process is also called \emph{order matchmaking}.
The second responsibility of electronic marketplaces is to execute trade between traders, as soon as a matching between an offer and a request is found.
The market operator either owns the items subject to trade (e.g., in forex markets), or acts as arbitrator in case of a dispute (e.g. AirBnb).

\subsection{Centralized Marketplaces}
To date, most electronic marketplaces are operated by a single market owner, responsible for defining and controlling the environment in which trade is executed~\cite{yarom2004decentralized}.
While it is a common practice to deploy marketplaces with a centralized authority and system architecture, it leads to a few deficiencies in general.
The first problem is that centralized systems tend to be less resistant against infrastructure failures and targeted attacks (e.g., a denial-of-service attack~\cite{mirkovic2004internet}).
Since digital assets are often stored on a single or a few servers, they pose an interesting target for attackers.
Second, the deployed business logic of (commercial) market operators is rarely open for inspection by traders.
The lack of audibility motivates market owners to exploit their prominent position.
For example, it is suspected that Uber manipulates ride prices with a dynamic pricing mechanism~\cite{chen2015peeking}.

The emergence of cryptocurrency exchanges, facilitating the trade of blockchain-based assets, further highlights issues underlying marketplaces under central control~\cite{moore2013beware}.
While a few of these exchanges process transactions worth millions of dollars in total daily, many market operators lack the knowledge or resources to quickly scale up their infrastructure to meet increasing demand.
More than once, the fragile infrastructure of centralized cryptocurrency exchanges has resulted in poor trading experiences, prolonged platform unavailability, and even the inability to withdraw assets from digital wallets by users~\cite{centralizedexchanges}.
The events surrounding the Mt. Gox cryptocurrency exchange in 2014, where hackers compromised Bitcoin worth around \$450 million, demonstrates that inadequate security measures can lead to irreversible reputational damage.

\subsection{Decentralized Marketplaces}
The shortcomings of centralized marketplaces motivate the deployment of blockchain-powered decentralized markets, or \emph{DEXes}.
On these DEXes, users can issue, manage and trade digital assets that are stored on a blockchain.
Furthermore, traders remain in full control of their assets, instead of having them managed by a single market operator.
DEXes leverage blockchain technology to transfer ownership of different types of assets at the same time, without the risk of these assets being compromised by an adversarial party.
Notable DEXes include Waves and BitShares, which obtained a market capitalization of \$80 million and \$83 million respectively at the time of writing~\cite{wavesplatform}~\cite{schuh2015bitshares}.
To date, there are more than 250 DEXes and over than 4.000 active traders across all DEXes~\cite{han2019optionality}.

A common characteristic of DEXes is that all orders created by traders are stored and processed on a blockchain.
To match compatible offers and requests, DEXes either execute matchmaking logic during transaction validation or they rely on a centralized server to conduct order matchmaking.
Although the market volume of DEXes is increasing, a DEX is limited to facilitate the trade of assets that are stored on its underlying blockchain.


\subsection{Asset Trading Between Different Blockchains}
Centralized cryptocurrency exchanges are able to trade assets stored on different blockchains through intermediate wallets owned by the market operator.
However, there are increasing trust issues around the companies operating these cryptocurrency exchanges~\cite{chohan2018problems}.
Although DEXes enable trust-less asset exchange without trusted third party, the trading is limited to assets managed on the blockchain used by a specific DEX.
The exchange of digital assets residing in two different blockchain platforms, e.g., exchanging Bitcoin and Ethereum assets, is a non-trivial problem.

The \emph{atomic swap} protocol addresses the problem of exchanging assets between different blockchains, without need for a trusted third party~\cite{herlihy2018atomic}.
Atomic swaps enable two parties to exchange blockchain-based assets in an atomic manner: the asset exchange either succeeds or fails for both parties at any given time.
The key technology underlying the atomic swap protocol is the Hashed Time-locked Contract, or HTLC~\cite{bowe2018hashed}.
This is a special type of blockchain transaction where a party must provide a pre-image of a cryptographic hash before some deadline, usually 24 hours, in order to claim his or her assets.
If this deadline is exceeded and a party did not provide the pre-image, the assets go back to the original owner.
Atomic swaps eliminate the \emph{counterparty risk} of losing assets to an adversarial trader while trading.
Counterparty risk is a serious concern in many electronic marketplaces that facilitate peer-to-peer trading~\cite{peters2016understanding}.

Although atomic swaps enable risk-free and cross-chain trading, we identify two deficiencies.
First, atomic swaps only work when trading assets between blockchains with support for specific programming constructs.
An atomic swap involving a blockchain platform without support for HTLCs is not possible.
Second, the atomic swap protocol is complicated to implement, and a single implementation mistake can result in the loss of funds.
We aim for a generic, cross-chain trading solution that does not rely on the availability of specialized transaction types.

\subsection{Problem Description}
\label{sec:problem_description}



In this work, we address the challenging problem of generic trade of digital assets between different blockchain platforms.
Due to the proliferation of blockchain-based assets, we believe that the availability of such a trading mechanism is becoming a necessity.
Based on the aforementioned deficiencies of existing centralized and decentralized marketplaces, we formulate three requirements for our cross-chain trading mechanism:

\begin{enumerate}
\item \textbf{Generic}. We require that our trading mechanism enables generic trade of assets between different blockchain platforms. In particular, the trade of assets with our mechanism should not be limited to a selected number of blockchain architectures with specific design requirements. We do not consider the atomic swap protocol an adequate solution for such a trading mechanism since it requires support for HTLCs.
\item \textbf{Decentralized}. We require that the architecture of our trading mechanism is decentralized, and that users themselves remain in control of the assets involved in a trade. More specifically, decentralization in the context of this work is achieved when there is not a single authority that controls the environment in which order matchmaking and trade execution take place. Trade should emerge and proceed through direct information exchange between traders and peer-to-peer payments.
\item \textbf{Limited counterparty risk}. During a trade, a counterparty might actively try to fraud the other party for economic benefit. In centralized marketplaces, counterparty risk is usually addressed by the (trusted) market operator which mediates the asset exchange between two parties. In decentralized marketplaces, counterparty risk is reduced since blockchain technology allows a trust-less transfer of assets. Our solution requires adequate measures to limit counterparty risk while trading.
\end{enumerate}

These requirements directly lead to the following research question: \emph{how can we devise a generic and decentralized mechanism to trade assets that are stored on different blockchains, with limited counterparty risk?}

\section{Solution Outline}
In this section, we outline our solution for generic, cross-chain asset trading.
During the following example, we assume that two traders, Alice and Bob, exchange blockchain-based assets.
Prior to the actual asset exchange, they have signed a contractual agreement containing all details of the upcoming trade.
These details include the negotiated exchange price and wallet addresses, and is further explained in Section~\ref{sec:protocol}.
The trade agreement between Alice and Bob is public and can be inspected by anyone.
Assume that Alice wants to sell her Bitcoin (BTC) to Bob, in return for some of his Ethereum gas (ETH).
A peer-to-peer asset exchange involves at least a transfer of BTC assets from the Bitcoin wallet of Alice to the Bitcoin wallet of Bob, and a transfer of ETH assets from the Ethereum wallet of Bob to the Ethereum wallet of Alice.

\begin{figure}[t]
	\centering
	\includegraphics[width=.7\linewidth]{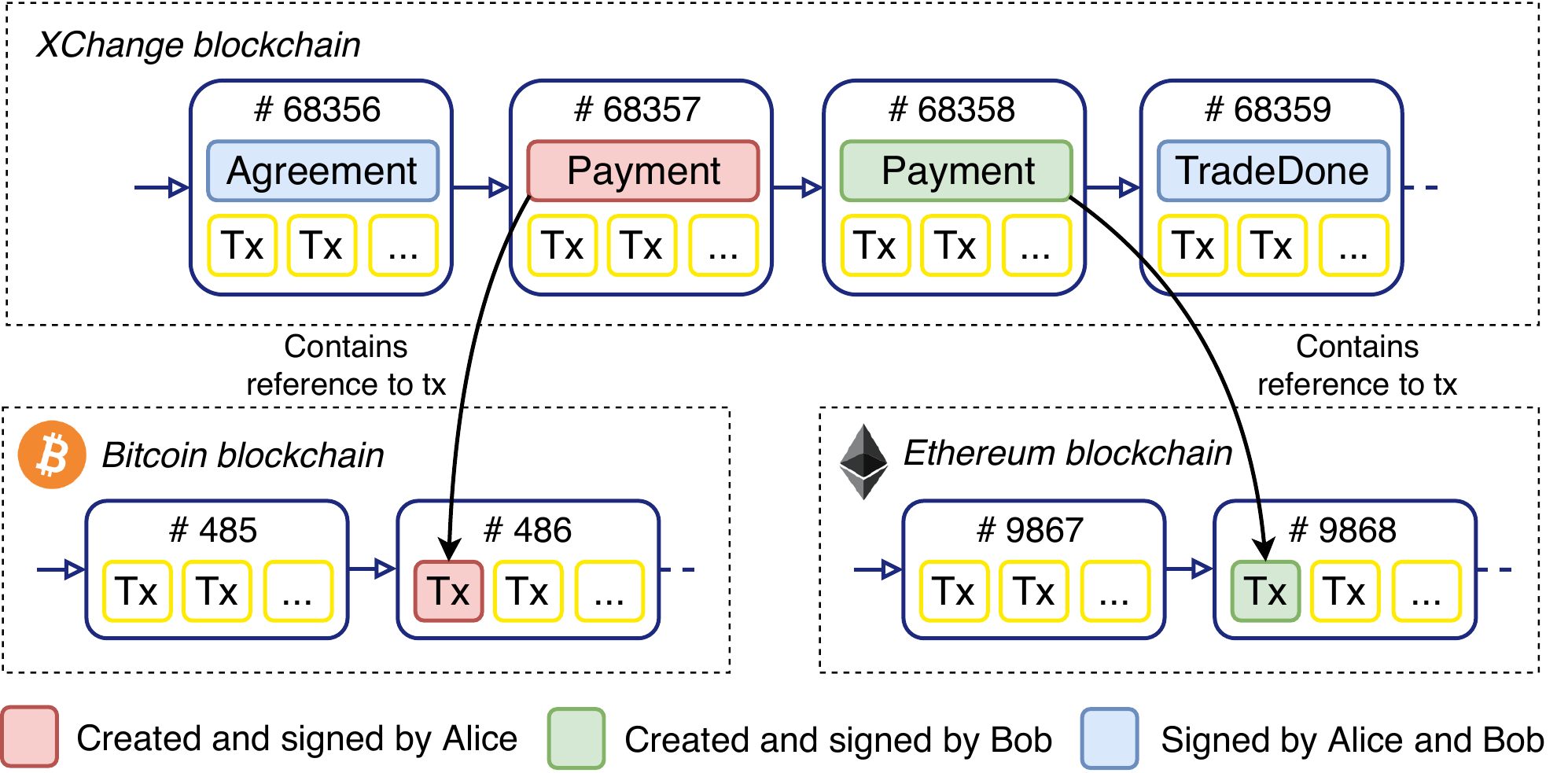}
	\caption{High-level overview of our \ModelName{} trading mechanism. In this example, Alice trades her Bitcoin for Bob's Ethereum tokens. Full trade specifications are stored on a separate blockchain.}
	\label{fig:xchange}
\end{figure}

\subsection{Naive Approach}
We first consider a naive approach to trade the BTC and ETH assets between Alice and Bob, without any trusted third party.
W.l.o.g., we assume that Alice initiates the trading process and has to initiate the trade by sending assets to Bob.
She does so by submitting a transaction to the Bitcoin blockchain which transfers BTC assets from her Bitcoin wallet to the Bitcoin wallet of Bob.
When this transaction is included on the blockchain, she notifies Bob about the conducted payment.
As soon as Bob verifies that the BTC arrived his Bitcoin wallet, he initiates a transfer of ETH from his Ethereum wallet to the Ethereum wallet of Alice.
Both these asset transfers are included as individual transactions in the Bitcoin and Ethereum blockchains, and are publicly visible to other interested parties.
Therefore, others can verify that all assets have been transferred between Alice and Bob and that both parties fulfilled all obligations as specified in the trade agreement.
The exchange of assets during a trade is also called \emph{settlement}.

A major problem with this trade procedure, however, is that Bob can commit \emph{counterparty fraud} by simply refusing to transfer his ETH to Alice as soon as Alice has transferred her BTC to him.
Likewise, if Bob would initiate the first payment to Alice, Bob is exposed to the counterparty risk of losing his ETH to Alice.
Centralized exchanges usually address this counterparty risk by active mediation during an asset exchange between two parties.
In this situation, both Alice and Bob transfer their BTC and ETH respectively to wallets owned by the market operator.
The market operator then transfers the BTC assets to the Bitcoin wallet of Bob and the ETH assets to the Ethereum wallet of Alice.

\subsection{Our Solution}
This work aims to limit counterparty risk during the asset exchange process by relying on \emph{accountability}.
The key idea of our trading mechanism, called \emph{\ModelName{}}, is to store full specifications of each trade on a separate blockchain.
Our approach is visualized in Figure~\ref{fig:xchange}, which shows (parts of) the Bitcoin and Ethereum blockchain in the lower part of the figure, and the separate blockchain used by \emph{\ModelName{}} in the upper part of the figure.
Technical requirements of the \ModelName{} blockchain are discussed in Section~\ref{sec:blockchain_accounting}, and for now we assume that this blockchain is capable of storing arbitrary data elements in a secure manner.

The trade between Alice and Bob now proceeds as follows: first, one of the parties publishes the dual-signed trade agreement on the \ModelName{} blockchain (in Figure~\ref{fig:xchange}, this agreement is included in the block with sequence number 68.356).
Next, the settlement process starts, and Alice and Bob initiate payments to each other.
W.l.o.g., assume Alice starts the trade by initiating a transaction to send her BTC to the Bitcoin wallet of Bob.
In Figure~\ref{fig:xchange}, this transaction is included in the block with sequence number 486 on the Bitcoin blockchain.
Next, Alice records this payment within a transaction on the \ModelName{} blockchain, which includes the identifier of the transfer transaction on the Bitcoin blockchain.
When Bob wants to verify that Alice has indeed transferred the BTC to his Bitcoin wallet, he can inspect the BTC transaction and check it for validity and finality.
When Bob has confirmed that Alice indeed sent the BTC to his Bitcoin wallet, he initiates a payment to Alice by submitting a transaction to the Ethereum blockchain.
In Figure~\ref{fig:xchange}, this transaction is included in the block with sequence number 9.868 on the Ethereum blockchain.
Next, Bob records this payment within a transaction on the \ModelName{} blockchain.
Alice now verifies whether Bob has sent the promised ETH assets to her Ethereum wallet.
If so, both parties explicitly publish a transaction on the \ModelName{} blockchain (e.g. block 68359 in Figure~\ref{fig:xchange}) to indicate that the trade has completed successfully.

Note that the above solution requires that the assets subjected to a trade can be publicly traced.
We believe this is not a major limitation since ownership change of a vast majority of blockchain-based assets is public information.
Furthermore, we assume that Alice and Bob can query the transaction in both the Bitcoin and Ethereum blockchain, e.g., by requesting the required information from a node that knows about all transactions on a specific blockchain (often referred to as a full node).
We consider privacy concerns beyond the scope of this work since they also apply to existing decentralized exchange solutions.

\subsection{Limiting Counterparty Risk}
\label{sec:limit_risk}
An important observation is that the above mechanism does not address counterparty risk when trading without trusted intermediary.
Our solution is still vulnerable when an adversary does not fulfil their obligations and steals assets from a counterparty.
Therefore, we extend our solution with two risk mitigation strategies:
\begin{enumerate}
\item \textbf{Limiting the number of outstanding trades with risky parties}. \ModelName{} relies on full accounting of trade specifications to limit the number of outstanding trades with risky parties.
A party is considered risky if it is currently responsible for the next payment during an ongoing trade.
By careful inspection of the information on the \ModelName{} blockchain, others can determine whether a specific party currently holds such a responsibility.
If so, the inspecting party should refuse to trade with the party that holds responsibility.
If all participants adhere to this strategy, an adversarial party can only commit counterparty fraud once.
Before others engage in a trade with a risky party, it must pass on the responsibility to a counterparty by initiating the next payment during an ongoing trade.
Nonetheless, a trader can still start a trade \textit{at its own risk} with a party that holds responsibility in a trade, e.g., with parties that a trader trusts.
\item \textbf{Incremental settlement}. In the example given above, a trade between Alice and Bob consists of two payments in total: one that transfers BTC assets and one that transfers ETH assets.
To further reduce counterparty risk, we can incrementally settle the trade by dividing each payment in $ n $ partial payments ($ n $ defaults to one).
The value of $ n $ should be included in the trade agreement.
For example, with $ n = 2 $, Alice first sends half of the agreed amount of BTC to Bob.
Bob then sends half of the agreed amount of ETH to Alice.
Incremental settlement decreases the economic benefit for adversarial parties when committing counterparty fraud, but prolongs the trade duration since more payments have to be made.
\end{enumerate}

The implementation of the \ModelName{} trading mechanism compromises of a five-phase trading protocol and a system architecture.
The formal specifications of the trading protocol are given in the next section whereas the system architecture of \ModelName{} is elaborated in Section~\ref{sec:architecture}.

\section{The \ModelName{} Trading Protocol} \label{sec:protocol}
In this section, we present the \ModelName{} trading protocol for generic asset trading between isolated blockchains.
The protocol describes how matchmakers match new orders submitted by traders, and coordinates the execution of a trade between trading parties.
The protocol design is based on five assumptions, which are stated below:

\begin{enumerate}
	\item \textbf{Strong identities}. The digital identity of each peer uniquely identifies a specific end-device in the network. 
	Well-established digital identities are necessary to prevent misbehaviours such as a Sybil Attack and a distributed denial-of-service attack~\cite{douceur2002sybil}\cite{Specht2004DistributedDO}.
	This is not an unrealistic assumption since many electronic marketplaces already impose some form of identity validation.
	Well-established digital identities are supported at the \textit{things layer} of IoT systems~\cite{wang2018privacy}.
	\item \textbf{Public-private keys}. Each peer in the network is in possession of a cryptographical key pair, consisting of a public and a private key. 
	The public key $ PK_p $ of a specific peer $ p $ is known to other peers and uniquely identifies this peer in the network. 
	Their private key, $ SK_p $, is used to digitally sign data such as outgoing messages and new orders. 
	We assume that adversarial parties cannot forge digital signatures.
	\item \textbf{Reliable network}. The transport layer of the network guarantees that messages eventually arrive their destinations if resubmitted often enough.
	\item \textbf{Availability}. Traders remain online while their orders are being matched and fulfilled.
	Since traders have an incentive to remain online to get their orders fulfilled, we believe this is a realistic assumption.
	\item \textbf{\ModelName{} blockchain}. Full trade specifications are stored on the \ModelName{} blockchain. This blockchain is denoted by $ \mathcal{B} $ during the protocol description.
\end{enumerate}

Besides storing trade specifications on $ \mathcal{B} $, \ModelName{} maintains an \emph{off-chain} network where traders negotiate trade and matchmakers bring traders together.
Matchmakers bundle unfulfilled orders they know about in their order books, and inform traders of potential trading partners.
An order book is an efficient data structure that stores orders and is optimized for finding the set of matching orders for an incoming order.
Every trader can act as a matchmaker within the \ModelName{} network at the cost of increased resource (to maintain an order book) and increased bandwidth usage (to process the orders sent by traders).
We denote the set of all available matchmakers in the network by $ \mathcal{M} $.
Our off-chain order book follows the decentralized approach adopted by the trading protocols AirSwap and 0x~\cite{airswap}\cite{warren20170x}.
Specifically, each trader sends a new order to one or more matchmakers.
The focus of this work is on the trading mechanism and we provide a full analysis of decentralized matchmaking in our ongoing work.\footnote{Omitted due to ongoing double-blind review, available on request.}

We now elaborate on each of the five phases during the \ModelName{} trading protocol.

\begin{lstlisting}[firstnumber=1,caption={Phase I: Order creation and matching.},label={alg:phase1},captionpos=b,numbers=left,float=t, tabsize=2, basicstyle=\footnotesize\ttfamily, mathescape=true, emph={while, if, then, do, Procedure, record, send, remove, insert, put, receive, in, foreach, ignore, else},emphstyle=\bf, frame=TB]
	Procedure create_order($ timeout, asset\_pair $):
		$ O $ $ \leftarrow $ $ (id, seq, timestamp, timeout, asset\_pair) $
		send Order($ O $) to $ m \subseteq \mathcal{M} $ 

	Procedure cancel_order($ O $):
		send CancelOrder($ O.id $, $ O.seq $) to $ m \subseteq \mathcal{M} $ 

	On receive of Order($ O $) or CancelOrder($ id $, $ seq $) message by peer $ p $:
		if message.type is Order and $ validate(O) $:
			insert $ O $ in order book
			$ find\_matches\_in\_order\_book(O) $
		else if message.type is CancelOrder:
			remove $ (id, seq) $ from order book

	Procedure find_matches_in_order_book(O)
		$ M \leftarrow $ matches(order_book, $ O $)
		foreach $ O_m $ in $ M $:
			send Match($ O, O_m $) to $ O.id $
	
	On receive of Match($ O, O_m $) message from $ m $ by peer $ p $:
		if $ p.has\_trade\_interest(O, O_m) $
			put Match($ O, O_m $) into $ match\_queues[O] $
		else:
			send RejectMatch($ O, O_m, reason $) to $ m $

	On Timeout of $ match\_queues[O] $:
		$ select\_trader\_from_\_queue(O) $
		
	Procedure select_trader_from_queue($ O $):
		$ counterparty $ $ \leftarrow $ $ select(match\_queues[O]) $
		$ verify\_trader(O, counterparty) $ # See phase II
		
	On receive of RejectMatch($ O, O_m, reason $) message:
		$ update\_order\_book(O, reason) $
		$ find\_matches\_in\_order\_book(O) $	
\end{lstlisting}

\begin{figure}[h]
	\centering
	\includegraphics[width=0.7\linewidth]{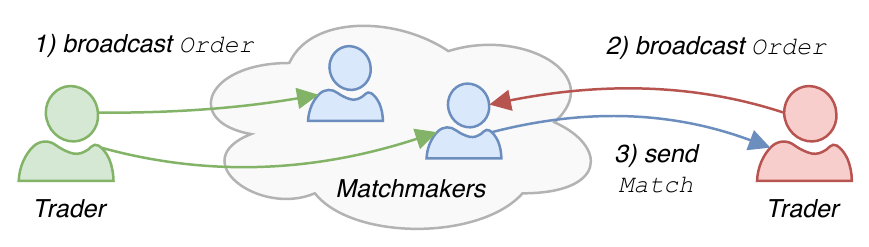}
	\caption{Phase I of the \ModelName{} trading protocol: traders (depicted in green and red) send new orders to matchmakers (shown in blue). Matchmakers notify traders about matches for their orders. }
	\label{fig:matching_protocol_1}
\end{figure}

\subsection{Phase I: Order Creation and Matching} \label{sec:phase_matching}

During the first phase of the \ModelName{} protocol, traders create new orders (offers and requests) and send these orders to matchmakers.
Matchmakers match incoming orders with existing orders in the order book, and notify order creators about trading opportunities.
Figure~\ref{fig:matching_protocol_1} visualizes the operations performed by traders and matchmakers during this phase, and Listing \ref{alg:phase1} lists the pseudo-code associated with this phase.

When a trader intends to exchange assets that it holds with assets that it wants to have, it creates a new order by calling the \texttt{create\_order} procedure (line 1, Listing \ref{alg:phase1}).
Each order $ O $ created by trader $ t $ is uniquely identified by ($ PK_t, O_{ind} $) where $ PK_t $ is the public key of $ t $ and $ O_{ind} $ is an ordinal positive integer indicating the order's position in the sequence of all orders of $ t $.
Furthermore, $ O $ contains a timestamp, indicating when the order has been created, a timeout, which specifies how long the order should remain valid, and a boolean value indicating whether the order is an offer or a request.
Inclusion of the timeout prevents orders from being open for an indefinite amount of time.
The content of $ O $ is digitally signed by $ t $ with $ SK_t $, to ensure non-repudiation and authenticity.
Finally, $ O $ contains specifications on the assets that $ t $ desires to buy or sell.
This information is provided as a tuple of assets, or an \emph{asset pair}.
If Alice wants to sell 1 BTC for 1 ETH, she creates an offer with asset pair (1 BTC, 1 ETH).
Similarly, if Bob wants to buy 1 ETH for 1 BTC, he creates a request with asset pair (1 BTC, 1 ETH).
The order creator includes the new order in a \MsgOrder{} message and disseminates it to a subset of all matchmakers (line 6), according to an order dissemination policy (see Section~\ref{subsec:overlay_logic}).

%

Traders are able to cancel an unfulfilled order $ O $ by sending a (signed) \MsgCancelOrder{} message with $ O_{ind} $ to matchmakers (line 6).
If a matchmaker receives such a message, it immediately deletes the cancelled order from its order book (line 13).

When a matchmaker $ m $ receives an \MsgOrder{} message (line 9), it extracts the order from the incoming message, validates its digital signature, and assesses correctness of the incoming order according to an order validation policy (see Section \ref{subsec:overlay_logic}).
If the incoming order is invalid or expired, $ m $ discards it and does not consider it for matching. 
If the incoming order is valid, it will be inserted in the order book of $ m $ and immediately matched with any existing orders with a mutual interest (line 16). 
If the matchmaker establishes a match, it informs the trader that created the incoming order about the trade potential.
In the situation that an order matches multiple orders, each matched order couple is considered as a different prospective trade. 
For instance, if an order of a trader $ t $ matches with $ n $ reciprocal orders, there are $ n $ potential trades in question, each of which includes $ t $ as one trading partner.
For each established match, the matchmaker sends a \MsgMatch{} message to the trader that creates the incoming order (line 18).
A \MsgMatch{} message contains full details on the matched orders.


The receiver of a \MsgMatch{} message is called the \emph{initiator}.
The initiator fulfils an important role if the match leads to a trade.
When trader $ t $ receives a \MsgMatch{} message from $ m $, it first checks whether its order is still valid, i.e., whether it still has an interest in trading the respective assets. 
If so, it remembers the matched order by adding it to a queue (line 22).
Otherwise, $ t $ sends a \MsgRejectMatch{} message back to $ m $ with a reason (line 24).
Upon reception of a \MsgRejectMatch{} by a matchmaker, it updates the status of the respective orders in their order book (line 34) and attempts to find new matches (line 35).

For each order, a trader $ t $ maintains a \emph{match priority queue} which contains the set of matched counterparties, nominated by matchmakers to $ t $. 
Each trader can have its own decision mechanism to select one trader amongst the others in the match priority queue.
An example decision mechanism would be to set a timer for the order and select the trader who has been nominated by the highest number of matchmakers when the timer expires. 
We motivate and elaborate on matching priority queues in Section \ref{subsec:match_priority_queue}.
We call the trader selected from the matching queue the \textit{counterparty} of a prospective trade.
Once $ t $ selects a counterparty, phase II of the \ModelName{} trading protocol starts and the initiator verifies the counterparty (line 31). 
Furthermore, the initiator reserves (\enquote{locks}) these assets to the counterparty which prevents the situation where it would incorrectly respond to a \MsgTrdProp{} message for a specific order while having an outstanding trade proposal for the same order.

\begin{figure}[h]
	\centering
	\includegraphics[width=0.6\linewidth]{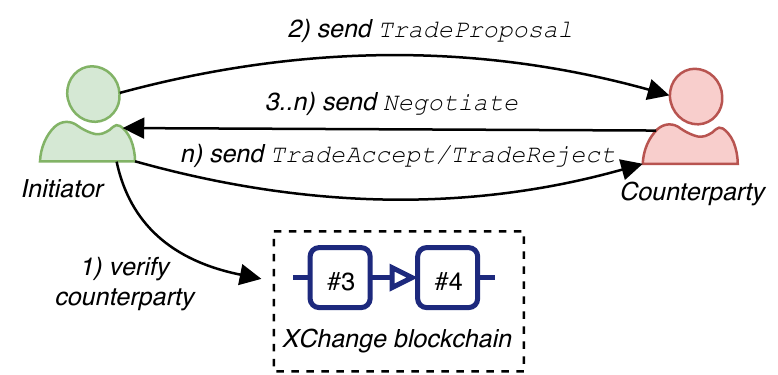}
	\caption{Phase II of the \ModelName{} protocol: two matched traders negotiate a trade.}
	\label{fig:matching_protocol_2}
\end{figure}

\subsection{Phase II: Trade Negotiation}
\label{sec:phase_negotiation}

The second phase of the \ModelName{} trading protocol is illustrated in Figure \ref{fig:matching_protocol_2} and formalized in Listing \ref{alg:phase2}.
During this phase, traders carry out off-chain trade negotiation with a sequence of network messages, without the involvement of matchmakers or other parties.
When an initiator has selected a counterparty for a potential trade, it first validates the counterparty by checking whether it holds responsibility during one of its ongoing trades (line 2, Listing~\ref{alg:phase2}).
Recall that a counterparty holds responsibility if it is their turn to initiate a payment to another party during an ongoing trade.
This can be verified by inspection of the transactions on $ \mathcal{B} $ that involves the counterparty.
Note that the presence of a \TRPayment{} transaction on $ \mathcal{B} $ is not enough to ensure that a party has actually transferred the promised assets.
Further inspection of the transaction identifier inside the \TRPayment{} transaction is necessary, which involves a transaction lookup in another blockchain.
An initiator following the \ModelName{} trading protocol will not engage in a trade with a counterparty that holds some responsibility.
This is a crucial property to address counterparty fraud in \ModelName{}.

\begin{lstlisting}[firstnumber=1,caption={Phase II: Trade negotiation.},label={alg:phase2},captionpos=b,float=t,numbers=left,tabsize=2, basicstyle=\footnotesize\ttfamily, mathescape=true, emph={while, if, then, do, Procedure, record, send, receive, in, foreach, ignore, else},emphstyle=\bf, frame=TB]
	Procedure verify_trader($ O, counterparty $):
		if $ check\_responsibilities(counterparty, \mathcal{B}) $
			send TradeProposal($ proposal $) to $ counterparty $
			remove $ counterparty $ from $ match\_queues[O] $
		else
			$ select\_trader\_from\_queue(O) $

	On receive of TradeProposal($ proposal $) or Negotiate($ proposal $) message by peer $ p $:
		$ result $ $ \leftarrow $ $evaluate(msg.proposal) $
		if $ result $ = ACCEPT:
			send TradeAccept($ proposal $) to $ msg.sender $
			if p is initiator:
				start_trade()
		if $ result $ = NEGOTIATE:
			send Negotiate($ proposal $) to $ msg.sender $
		if $ result $ = REJECT:
			send TradeReject($ proposal $) to $ msg.sender $
			if p is initiator:
				send RejectMatch($ proposal.O, proposal.O_m, reason $) to $ m \subseteq \mathcal{M} $
				$ select\_trader\_from\_queue(O) $
		
	On receive of TradeReject($ proposal $) message:
		if initiator:
			send RejectMatch($ proposal.O, proposal.O_m, reason $) to $ m \subseteq \mathcal{M} $
			$ select\_trader\_from\_queue(O) $
	
	On receive of TradeAccept($ proposal $) message by peer $ p $:
		if p is initiator:
			$ construct\_trade\_agreement(proposal)	$  # See phase III
\end{lstlisting}

If the initiator wishes to trade with a counterparty after the verification of their responsibilities, it sends a \MsgTrdProp{} message (line 3).
This message contains the order of the initiator, the identifier of the (matched) order of the counterparty, and a trade proposal.
When the counterparty receives a \MsgTrdProp{}, it has a chance to negotiate by sending \MsgTrdNegotiate{} message back to the initiator (line 15).
This might happen when the counterparty already fulfilled a part of their order and the initiator proposes to trade more assets than the counterparty is willing to.
This situation could occur if the matchmaker has outdated information on the status of the counterparty's order.

Trade negotiation may continue by an exchange of \MsgTrdNegotiate{} messages between the initiator and the counterparty, until one of the two sides eventually accept or reject the last negotiation offer by a \MsgTrdAccept{} or \MsgTrdReject{} message, respectively.
If one of the parties has sent a \MsgTrdAccept{} message, the initiator immediately starts the preparation of trade, entering phase III of the trading protocol (line 29). 
In the situation where one side has sent \MsgTrdReject{} message, the initiator will send a \MsgRejectMatch{} message to the matchmaker to inform them about the failed negotiation, and select a new counterparty from the match priority queue, if available (lines 19 and 25).
The outcome of the negotiation phase is an asset pair which specifies which and how many assets will be exchange between the involved traders.

\begin{lstlisting}[firstnumber=1,caption={Phase III: Constructing a trade agreement between an initiator and a counterparty.},label={alg:phase3},captionpos=b,float=t,numbers=left,tabsize=2, basicstyle=\footnotesize\ttfamily, mathescape=true, emph={while, if, then, do, Procedure, record, send, receive, in, foreach, ignore, else},emphstyle=\bf, frame=TB]
	Procedure construct_trade_agreement($ proposal $)
		$ A_{partial} \leftarrow proposal.to\_agreement() $
		$ A_{partial}.add\_wallet\_info() $
		send PartialAgreement($ A_{partial} $) to $ proposal.counterparty $

	On receive of PartialAgreement($ A_{partial} $) message:
		$ A \leftarrow A. add\_wallet\_info() $
		send Agreement to $ A_{partial}.initiator $

	On receive of Agreement($ A $) message:
		$ \mathcal{B}.publish(A) $
		$ trade \leftarrow A.to\_trade() $
		insert $ trade $ in $ trades $
		$ send\_payment(trade) $ # See phase IV
\end{lstlisting}

\begin{figure}[h]
	\centering
	\includegraphics[width=0.6\linewidth]{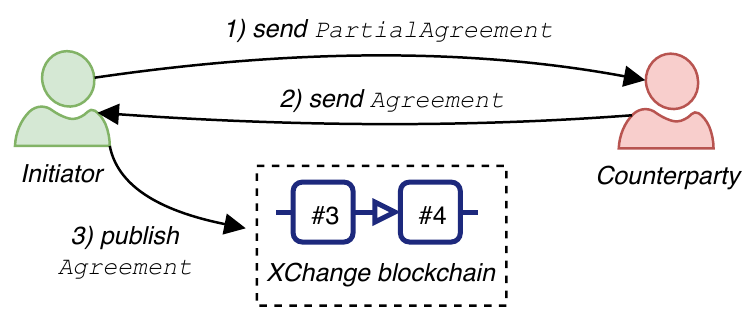}
	\caption{Phase III of the \ModelName{} protocol: the initiator and counterparty construct a trade agreement, which is published on the \ModelName{} blockchain by the initiator.}
	\label{fig:matching_protocol_3}
\end{figure}

\subsection{Phase III: Constructing a Trade Agreement} \label{sec:phase_agreement}

In the third phase of the trading protocol, traders enter into an irrevocable trade agreement and publish this agreement on the \ModelName{} blockchain.
This phase is visualized in Figure \ref{fig:matching_protocol_3} and formalized in Listing \ref{alg:phase3}.

When one of the parties accept the trade proposal, the initiator starts the third phase of the trading protocol by constructing a partial trade agreement, including local information regarding the upcoming trade (line 1, Listing~\ref{alg:phase3}).
This information includes the wallet address(es) of the initiator where the counterparty is expected to send assets to, the public keys (identifiers) of both traders, and the asset pair that is negotiated during the negotiation phase.
The trade agreement also contains information on whether incremental settlement is used, and if so, the number of expected payments originating from each side.
The initiator embeds this information in a \MsgPartialAgreement{} message, signs it, and sends to the counterparty (line 4).
When the counterparty receives the \MsgPartialAgreement{} message, it verifies the included information and creates an \MsgAgreement{} message which includes the wallet address(es) of the counterparty (line 6-8).
The counterparty signs the \MsgAgreement{} message and sends it back to the initiator, who then publishes the dual-signed trade agreement on $ \mathcal{B} $ (line 11).
Each trade agreement contains a \emph{publication deadline}.
If the trade agreement is published on $ \mathcal{B} $ after the publication deadline passed, the trade agreement is considered invalid.
Inclusion of this deadline addresses the attack scenario where an initiator purposefully withholds a dual-signed trade agreement to avoid taking responsibility in the trade specified by the agreement.


\begin{lstlisting}[firstnumber=1,caption={Phases IV: Execution of a trade.},label={alg:phase4},captionpos=b,numbers=left,float=t, tabsize=2, basicstyle=\footnotesize\ttfamily, mathescape=true, emph={while, if, then, do, Procedure, record, send, initiate, receive, in, foreach, ignore, else},emphstyle=\bf, frame=TB]
	Procedure send_payment(trade)
		$ payment \leftarrow trade.get\_next\_payment() $
		$ payment.txid \leftarrow conduct\_payment(payment) $ # Transfer assets on blockchain
		$ \mathcal{B}.publish(payment) $
		send Payment($ payment $) to $ trade.counterparty $
		if $ trade.complete() $:
			$ construct\_trade\_done(trade) $ # See phase V
		
	On receive of Payment($ payment $) message by peer $ p $:
		$ trade \leftarrow payment.trade\_id $
		$ valid \leftarrow payment.validate() $ # Is the payment initiated by the other side?
		if $ valid $ and not $ trade.complete() $:
			$ send\_payment(trade) $
		elif $ valid $ and $ trade.complete() $ and is initiator:
			$ construct\_trade\_done(trade) $ # See phase V
\end{lstlisting}

\begin{figure}[h]
	\centering
	\includegraphics[width=0.7\linewidth]{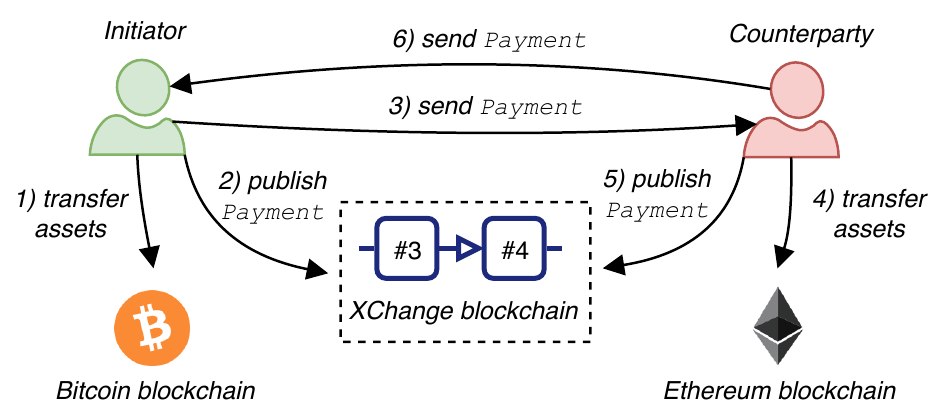}
	\caption{Phase IV of the \ModelName{} protocol: The trade is executed by a sequence of payments. In this figure, Bitcoin are exchanged with Ethereum assets, and the trade consists of two payments.}
	\label{fig:matching_protocol_4}
\end{figure}

\subsection{Phase IV: Trade Execution} \label{sec:phase_execution}

During the fourth phase of the \ModelName{} trading protocol, parties execute the trade and exchange assets in accordance with the trade agreement.
This phase is visualized in Figure~\ref{fig:matching_protocol_4} and the pseudo-code is provided in Listing \ref{alg:phase4}.

A trade usually consists of two payments (asset transfers), but more payments are required if the parties agreed on using incremental settlement.
The initiator always has the responsibility to make the first payment in a trade.
The initiator initiates a payment $ p $ by initiating an asset transfer on the blockchain that hosts the assets subject to trade (line 3, Listing~\ref{alg:phase4}).
This should result in a transaction identifier associated with the payment, say $ tx_p $.
The initiator then publishes the payment information on $ \mathcal{B} $ which includes $ tx_p $.
The \TRPayment{} transaction contains sufficient information for the counterparty to assess whether the initiator actually made the payment.
The transaction also includes a reference to the trade agreement.
The initiator then sends a \MsgPayment{} message to the counterparty to inform them about the payment (line 5).
The counterparty now polls its asset wallet to verify whether the promised assets have arrived (line 11).
If so, the counterparty initiates a payment back to the initiator.
This process repeats until all payments have been made and all assets have been transferred between the initiator and the counterparty.
At that point, the initiator finalizes the trade during the last phase of the trading protocol (lines 7 and 15).



\begin{figure}[h]
	\centering
	\includegraphics[width=0.7\linewidth]{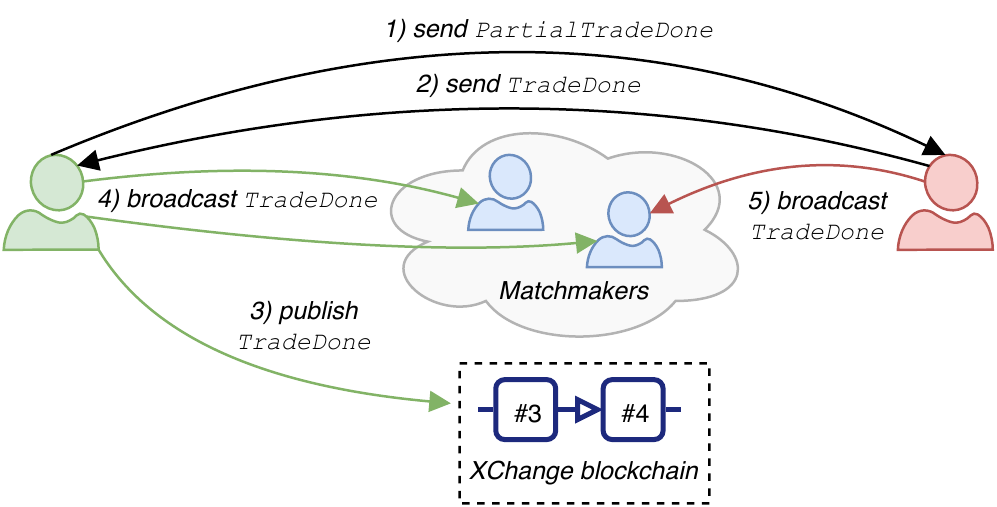}
	\caption{Phase V of the \ModelName{} protocol: The trade is finalized and matchmakers are informed about the conducted trade.}
	\label{fig:matching_protocol_5}
\end{figure}

\subsection{Phase V: Trade Finalization}
\label{sec:phase_finalization}
Once both parties have completed all payments and have created the \TRPayment{} transaction on $ \mathcal{B} $, they finalize the trade using a dual-signed transaction.
\ModelName{} explicitly finalizes, or \enquote{closes} a trade between two parties in order for others to quickly verify that a trade has been completed successfully, without having to verify individual \TRPayment{} transactions.
This process is visualized in Figure~\ref{fig:matching_protocol_5} and the associated pseudo-code is given in Listing~\ref{alg:phase5}. 
The initiator of a trade first sends a signed \MsgPartialTradeDone{} message to the counterparty (line 3, Listing~\ref{alg:phase5}).
This message is sent to the counterparty and contains the identifiers of the \TRAgreement{} and all \TRPayment{} transactions included on $ \mathcal{B} $.
Upon reception of this message by the counterparty, it signs the partial \MsgPartialTradeDone{} (line 6) and informs the matchmakers about the completed trade (line 8).
The counterparty then sends a \MsgTradeDone{} message back to the initiator (line 7).
When the initiator receives this message, it publishes the dual-signed trade finalization in a \TRTradeDone{} message on $ \mathcal{B} $ (line 11) and informs matchmakers about the completed trade (line 12).
When a matchmaker receives a \MsgTradeDone{} message, it updates the status of the orders of the initiator and counterparty in their order book.

\begin{lstlisting}[firstnumber=1,caption={Phases V: Finalization of a trade.},label={alg:phase5},captionpos=b,numbers=left,float=t, tabsize=2, basicstyle=\footnotesize\ttfamily, mathescape=true, emph={while, if, then, do, Procedure, record, send, initiate, receive, in, foreach, ignore, else},emphstyle=\bf, frame=TB]
	Procedure construct_trade_done(trade)
		$ D_{partial} \leftarrow trade.to\_trade\_done() $
		send PartialTradeDone($ D_{partial} $) to $ trade.counterparty $
		
	On receive of PartialTradeDone($ D_{partial} $) message by peer $ p $:
		$ D \leftarrow p.sign(D_{partial}) $
		send TradeDone(D) to $ m \subseteq \mathcal{M} $
		send TradeDone(D) to $ trade.initiator $
	
	On receive of TradeDone($ D $) message:
		$ \mathcal{B}.publish(D) $
		send TradeDone(D) to $ m \subseteq \mathcal{M} $
	
	On receive of TradeDone($ D $) message by matchmaker $ m $:
		$ update\_order\_book(D) $
\end{lstlisting}

\subsection{Security Analysis} \label{sec:analysis}
We now present a basic security analysis of the \ModelName{} trading protocol and analyse the consequences of an adversarial trader that diverts from the protocol.
We leave a discussion of the security guarantees of the blockchain used by \ModelName{} to other work since these guarantees highly depend on the specifications of $ \mathcal{B} $.
The following analysis highlights potential attacks during the \ModelName{} trading protocol where an adversarial party attempts to purposefully compromise assets or stale  the activity of ongoing trades.

\textbf{Order creation and matching (phase I)}.
During the first phase, matchmakers may intentionally match orders according to their own desires.
For example, a matchmaker could give preferential treatment to orders made by specific traders, and report better matches for their orders.
However, since a trader disseminates new orders to multiple matchmakers simultaneously, there is a high probability that at least one matchmaker honestly reports matches back to the trader.
This probability increases if the order is sent to more matchmakers after its creation.

\textbf{Trade negotiation (phase II)}.
During the second phase, a trader first verifies the responsibilities of a counterparty before sending out a \MsgTrdProp{} message.
This is a key mechanism to limit counterparty fraud.
If $ t $ skips this verification, it opens a window of opportunity for the counterparty to commit fraud with multiple parties simultaneously, including $ t $.
If $ t $ acts rational, it is in their best interest to perform this verification by inspecting the outstanding trades of the counterparty to lower the risk of losing assets.

During the trade negotiation, any party can leave ongoing negotiations without repercussions by sending a \MsgTrdReject{} message to the other side.
This includes the situation where one of the parties does not respond with a \MsgTrdNegotiate{} or \MsgTrdAccept{} message in a timely manner.
The party that left the negotiations can now start new negotiations with another potential trading partner.

\textbf{Constructing a trade agreement (phase III)}.
While constructing a trade agreement, the initiator does have an incentive to defer the publication of a dual-signed trade agreement on $ \mathcal{B} $.
Specifically, as soon as the trade agreement is published on $ \mathcal{B} $, the initiator takes on responsibility during the agreed trade.
Inspection of the responsibilities of the initiator by others will reveal his position in the agreed trade, and other parties will refuse to trade with the initiator.
To ensure that the initiator publishes the agreement in a timely manner on $ \mathcal{B} $, the trade agreement contains a publication deadline (see Section~\ref{sec:phase_agreement}).
A trade agreement published after its publication deadline by the initiator is considered invalid by others, and the two parties that signed the expired agreement will not continue to the next phase of the trading protocol.
The publication deadline also forces the counterparty to respond to an incoming \MsgPartialAgreement{} message in a timely manner.

\textbf{Trade execution (phase IV)}.
During trade execution, a trader $ t $ might refrain from initiating a payment to the counterparty after having received assets from the counterparty.
However, other rational traders will not engage in a trade with $ t $ until $ t $ has passed on its responsibility.
Since \ModelName{} assumes that the identity of each trader is verified, $ t $ cannot easily re-enter the network under a new identity when it has committed fraud.
Also note that rational traders are not likely to refrain from publishing their \TRPayment{} transaction on $ \mathcal{B} $, since doing so will pass on their responsibility during an ongoing trade.

\textbf{Trade finalization (phase V)}.
When a trade between two parties enters the finalization phase (see Section~\ref{sec:phase_finalization}), all required payments have been made and all required \TRPayment{} transactions have been published on $ \mathcal{B} $.
Even if one of the parties actively delays the finalization phase, others can still assess that the trade has been completed by inspection of the last \TRPayment{} transaction.

\section{The \ModelName{} System Architecture}
\label{sec:architecture}
\begin{figure*}[t]
	\centering
	\includegraphics[width=\linewidth]{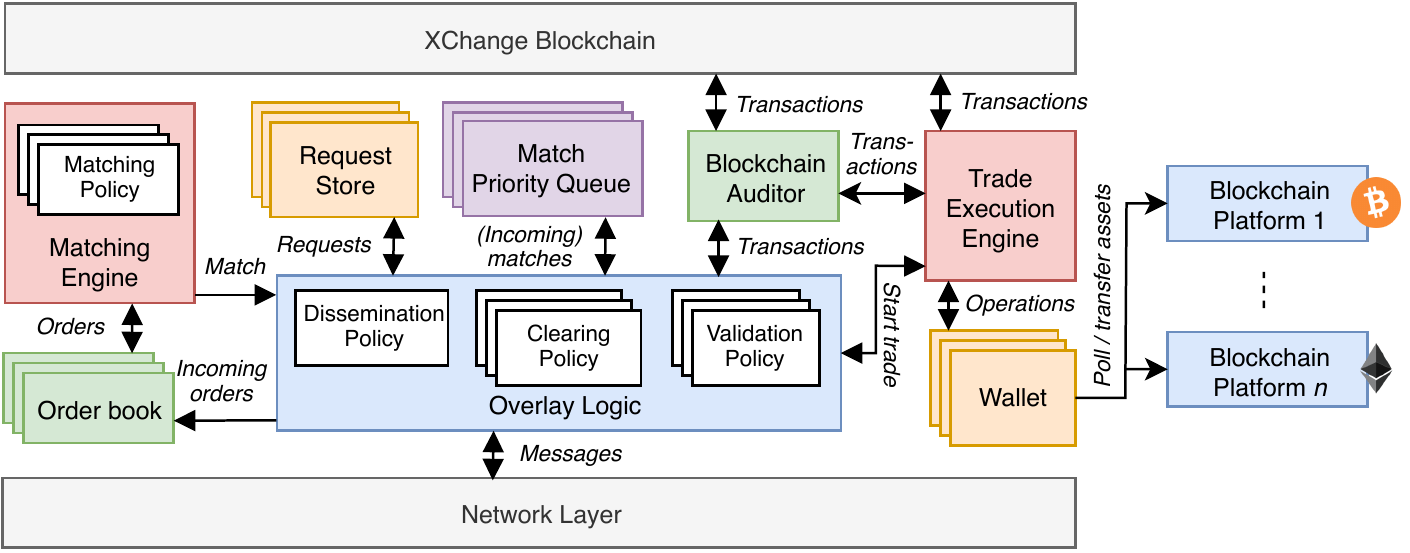}
	\caption{The system architecture of \ModelName{}, our mechanism for generic trade at scale. Arrows between two components indicate interaction or data exchange.}
	\label{fig:mechanism_architecture}
\end{figure*}

We now present the \ModelName{} system architecture, visualized in Figure \ref{fig:mechanism_architecture}.
Each active peer in the \ModelName{} network runs a single implementation of this system architecture as a shared library.
We briefly explain each component and highlight how interaction proceeds between components.

\subsection{Network Layer}
\label{subsec:network_layer}
The network layer passes incoming messages to the overlay logic and routes outgoing messages to their intended destination.
This layer can be implemented using any networking library with support for peer-to-peer communication (for example, libp2p\footnote{https://libp2p.io}).
The \ModelName{} trading protocol does not make assumptions about the topology of the network layer, except for the fact that each peer should know the network address of some matchmakers that can match their orders.
In our implementation, we use a bootstrapping mechanism where new peers in the network immediately attempt to discover some available matchmakers.
The networking library manages this bootstrapping process.
Attacks targeted at the network layer (e.g., the Eclipse Attack~\cite{singh2006eclipse}) are beyond the scope of this work.
The Sybil Attack is addressed by the requirement for strong identities, as discussed in Section \ref{sec:protocol}.

\subsection{Overlay Logic}
\label{subsec:overlay_logic}
The \ModelName{} overlay logic (depicted in blue in~Figure \ref{fig:mechanism_architecture}) reacts to incoming network messages and optionally passes the information in these messages to the relevant components.
For example, an incoming order (embedded in an \MsgOrder{} message) is passed to the order book and stored there.
Similarly, the information in an incoming \MsgMatch{} message is passed to the match priority queue.
The overlay logic interacts with the trade execution engine when a new trade is about to start, e.g., when receiving a \MsgTrdAccept{} message.
It also passes information about an incoming \MsgPayment{} message to the trade execution engine.
The overlay logic defines three different policy types, for order dissemination, order validation, and (trade) clearing.
These policies are explained next.

\textbf{Dissemination policy}.
The dissemination policy specifies how a new order is disseminated to matchmakers.
It contains information to which matchmakers a new order should be send.
The number of matchmakers that receive a new order is also called the \emph{order fanout}.
Sending a new order to more matchmakers potentially speeds up the matching of this order, but it also increases the load on matchmakers since they have to process more incoming \MsgOrder{} messages.

\textbf{Clearing Policies}.
The clearing policies predicate whether a specific peer should send out a trade proposal to another trader, or accept an incoming trade proposal that it received (during phase II of the \ModelName{} trading protocol, see Section \ref{sec:phase_negotiation}).
\ModelName{} enables programmers to define distinct and multiple clearing policies for orders with differing specifications.
For example, a clearing policy could specify that a trader can only exchange assets with other traders that are within geographic proximity.
Another clearing policy could implement trading restrictions governed by regional or local legislation.
We have implemented the strategy that a trader only interacts with traders to do not hold any responsibility during trade as a clearing policy.
This policy is enabled by default.

\textbf{Validation Policies}.
The validation policies implement validation rules for new incoming orders.
Similar to the clearing policies, multiple validation policies can be applied to orders with differing specifications.
An order validation policy often takes into consideration the information in each order and determines whether this attribute data is valid with respect to a specific domain
(e.g.\ whether the quantity of stocks is non-negative).
The overlay logic immediately discards orders that are classified as invalid by any applied validation policy.

\subsection{\ModelName{} Blockchain}
The \ModelName{} trading protocol assumes the availability of a blockchain on which full trade specifications are stored.
The popularity of the Bitcoin cryptocurrency bootstrapped the development of an extensive range of blockchain platforms with varying security and performance guarantees.
Most of these ledgers are designed to securely manage and transfer cryptocurrencies or digital tokens.
The distributed ledger used by \ModelName{} does not necessarily need the required security level of managing digital currencies.
Instead, we require a blockchain that merely allows for the tamper-proof storage of trade specifications.
We explore suitable blockchains for the \ModelName{} trading mechanism in Section \ref{sec:implementation}.

\subsection{Order Book}
\label{subsec:order_book}
An order book is a data structure that stores (active) orders.
It lists the specific assets that are being bought or sold within the market and provides traders with a convenient view of the current supply and demand.
Depending on the trading environment, orders in the order book often contains the identity of its creator, and sometimes the reputation scores of the order creator (for instance, eBay shows the trustworthiness of merchants).
When presenting an order book to traders, orders are usually sorted on one or more attributes, e.g., based on their price.
For orders with pricing information attached, the order book usually lists the \enquote{best} orders at the top.

In \ModelName{}, each matchmaker can operate multiple order books for different order types (coloured green in Figure~\ref{fig:mechanism_architecture}).
Specifically, orders exchanging the same asset pair are stored in the same order book.
An \emph{asset pair} specifies the pair of assets that are being traded in an order book and uniquely identifies each order book.
For example, an order book identified by the BTC/ETH asset pair contains orders that buy or sell Bitcoin and Ethereum gas.
Maintaining distinct order books at the same time enables order management with a single infrastructure across different trading domains.

\subsection{Matching Engine}
\label{subsec:local_order_book}
The matching engine (coloured red in Figure~\ref{fig:mechanism_architecture}) matches incoming new orders with existing orders in order books.
Marketplaces operated by a central authority often deploy an in-house order matching engine, fully controlled by the market operator.
DEXes sometimes integrate the order matching process in the transaction validation process.
In \ModelName{}, each matchmaker operates a matching engine and immediately attempts to match new incoming orders.
Established matches for an incoming order are passed to the overlay logic, embedded in a \MsgMatch{} message and sent to the order creator.

The matching engine can contain multiple \emph{matching policies}.
A matching policy predicates whether an offer and request match, and is applied to a single type of order book.
It takes a single offer and request as input and determines whether these orders match, based on their specifications.
In the context of cryptocurrency trading, whether two orders match depends on the asset price contained in the order.
The most common matching strategy in financial markets is the \emph{price-time matching strategy}, where orders are first matched based on their price, and then based on order creation time in case of a tie-breaker (older orders are prioritized).
\ModelName{} allows programmers to define custom matching policies.

\subsection{Request Stores}
To correctly process incoming \MsgTrdAccept{} and \MsgTrdReject{} messages during trade negotiation (phase II of the \ModelName{} trading protocol, see Section~\ref{sec:phase_negotiation}), \ModelName{} stores the state of outgoing \MsgTrdProp{} and \MsgTrdNegotiate{} messages.
The state of outgoing messages is stored in distinct \emph{request stores} (coloured yellow in Figure~\ref{fig:mechanism_architecture}).
For each outgoing message that has a state attached, a unique identifier is generated, a new request store containing this identifier is created and the generated identifier is appended to the outgoing message.
Traders that receive a message with this identifier are required to include the same identifier in their response message.
Incoming response messages with an unknown identifier are discarded and not processed further.
Each request store can have an optional timeout, indicating the duration after which the request store times out.
On timeout of a request store, it is deleted, and the overlay logic is notified about this event.

\subsection{Match Priority Queue (MPQ)} 
\label{subsec:match_priority_queue}

After creating a new order, a trader can receive multiple \MsgMatch{} messages from matchmakers within a short period of time.
A basic approach is to process incoming \MsgMatch{} messages on a first-in-first-out (FIFO) basis.
However, this approach leaves the mechanism vulnerable to an attack where a malicious matchmaker is the first to send a specific \MsgMatch{} message with low quality to a trader, which is then immediately processed by this trader.
The trader might have received a better match from an honest matchmaker if it would have waited longer before processing incoming \MsgMatch{} messages.

This issue is addressed as follows: incoming \MsgMatch{} messages for a specific order are first collected and stored in MPQ, coloured purple in Figure \ref{fig:mechanism_architecture}.
When a trader receives the first \MsgMatch{} message for their specific order $ O_1 $, it creates a new match priority queue for $ O_1 $.
Each entry in the MPQ of order $ O_1 $ is a two-dimensional tuple $ (r, O_2) $ where $ r $ indicates the number of retries, and $ O_2 $ is the order that matches with $ O_1 $.
Removing items from the matching queue is first based on $ r $ (item with a lower value of $ r $ have a higher priority) and then based on the quality of the match (better matches have a higher priority).
The number of retries $ r $ in each entry $ (r, O_2) $ indicates how many times we tried to negotiate a trade with the creator of $ O_2 $.
This value is initially zero and incremented by one each time a trade proposal is denied for the reason that a counterparty does not have any unreserved assets.
Specifically, when a trader receives a \MsgTrdReject{} message for this reason from the creator of $ O_2 $ during trade negotiation for his own order $ O_1 $, it adds the entry $ (r + 1, O_2) $ to the MPQ of $ O_1 $.

Before trade negotiation with a specific trader starts, incoming matches for each order are stored in MPQ for some duration $ W_m $, also called the \emph{match window}.
The value of $ W_m $ should be carefully considered: a high value of $ W_m $ increases the probability of receiving good matches but adds to the order completion time since a trader has to wait longer before sending out trade proposals.
On the other hand, decreasing the match window might lead to missing better matches.
When the match window expires, a trader removes an entry $ (r, O) $ from the MPQ, verifies the counterparty and sends out a \texttt{ProposeTrade} message to the trader that created $ O $ when the the trader wishes to trade with the counterparty (also see Section~\ref{sec:phase_negotiation}).
Only when trade negotiation for an order fails or succeeds, the next entry is removed from the appropriate MPQ if there is still trading opportunity for order $ O $.
If $ O $ is fulfilled, the MPQ associated with $ O $ is deleted.

\subsection{Blockchain Auditor}
\label{subsec:ledger_auditor}
The blockchain auditor (coloured green in Figure~\ref{fig:mechanism_architecture}) inspects the transactions that are stored on the \ModelName{} blockchain and reports this information to the trade execution engine or the overlay logic.
Conceptually, it converts the information in transactions stored on the distributed ledger to domain-specific knowledge that can be used by different components.
For example, prior to trade negotiation between two traders, the overlay logic invokes the blockchain auditor to determine whether a trader $ t $ holds responsibility in an outstanding trade with another party (which is determined by inspecting the transactions which involve $ t $).

\subsection{Trade Execution Engine}
\label{subsec:trade_execution_engine}
The trade execution engine manages and controls the trading process.
It tracks the current state of a specific trade and interacts with the \ModelName{} blockchain to store trade agreement, payment specifications and trade finalizations.

\subsection{Wallets}
\label{subsec:wallets}
\ModelName{} organizes different types of assets within wallets.
These wallets organize the information provided by external blockchain platforms (e.g., Bitcoin or Ethereum) in a generic manner.
Wallets are designed to store information on any digital asset such as cryptocurrencies or digital tokens.
Wallets expose functionality to transfer available assets to others, to query the current balance and to fetch past transactions.
Each wallet is uniquely identified by a \emph{wallet address}, usually a public key.
The addresses of wallets where assets should be sent to are included in a \TRAgreement{} transaction on the \ModelName{} blockchain.

\subsection{External Blockchain Platforms}
As outlined in the protocol description, the \ModelName{} blockchain stores full trade specifications.
\ModelName{} interacts with external blockchain platforms, such as the Bitcoin or Ethereum platform, to transfer assets to the counterparty.
Specifically, wallets invoke operations to query external blockchain platforms whether assets have been received, or to transfer assets.

\section{Blockchain-based Accounting of Trade Specifications}
\label{sec:blockchain_accounting}
The \ModelName{} trading protocol (see Section~\ref{sec:protocol}) requires a blockchain to store full trade specifications.
The popularity of blockchain technology has resulted in much interest to design and deploy distributed ledgers with different scalability and security guarantees.
In this section, we review different blockchain organizations and show which structures synergies with the \ModelName{} trading protocol.


\subsection{Traditional Blockchain Ledgers}

In most traditional blockchain ledgers, there is a global consensus mechanism that periodically determines a leader amongst all consensus participants, or \emph{miners}.
This leader is then allowed to extend the blockchain with exactly one block.
In the Bitcoin network, a new block is appended roughly every ten minutes~\cite{nakamoto2008bitcoin}.
For Bitcoin and Ethereum, each miner in the network attempts to solve a computational puzzle and the first miner to solve the puzzle can append a new block to the blockchain ledger (Proof-of-Work).
Verifying the correctness of a proposed solution is computationally efficient and can be performed in $ O(1) $ time.
The difficulty of the computational puzzle grows with the overall computing power of the network.
Other consensus mechanisms might select a leader based on the miner's stake in the network (Proof-of-Stake) or by election of a committee that produces blocks (Delegated Proof-of-Stake)~\cite{bentov2014proof}.
For \ModelName{}, and even within Internet-of-Things applications in general, we consider the usage of blockchain ledgers secured by global consensus impractical for the following two reasons.\footnote{Still, the \ModelName{} trading protocol can leverage any distributed ledger that is secured by global consensus and has accounting capabilities.}

\begin{enumerate}
	\item \textbf{Inefficient.} Reaching a global, network-wide consensus is often an expensive process in terms of CPU power, bandwidth usage and storage requirements.
	It negatively impacts the throughput of a blockchain in terms of finalized transactions per second.
	The theoretical throughput of a blockchain ledger is often bounded by a constant value.
	Gervais et al. found that global consensus mechanisms based on computing power (Proof-of-Work) can achieve up to around 60 transactions per second before security issues arise~\cite{gervais2016security}.
	Considering the throughput needed to process payments at a global scale, many blockchains are by far not scalable enough to provide viable alternatives for financial use-cases such as asset marketplaces.
	\item \textbf{Transaction fees.} To incentive participation in the consensus mechanism, miners are rewarded by transaction fees.
	These transaction fees are covered by users that initiate a transaction.
	Depending on the blockchain used by \ModelName{}, the transaction fees required to get the transactions included during a trade might be relatively high, especially if a trader only desires to exchange a few assets with relatively low value.
	Transaction fees potentially make asset trading with \ModelName{} a costly process.
\end{enumerate}




\subsection{DAG-based Blockchains}
As an attempt to increase the throughput of distributed ledgers, various platforms reorganize the structure of their blockchain and maintain a Directed Acyclic Graph (DAG) structure instead~\cite{sompolinsky2018phantom}.
Notable applications that deploy a DAG structure include IOTA, Nano and Dagcoin~\cite{popov2016tangle}~\cite{lerner2015dagcoin}.
Its distinguishing property is that each transaction can be referenced by multiple other ones and as such, it forms a DAG of transactions.
Transactions in a DAG are linked, and each transaction verifies one or more prior transactions.

Since DAG-based ledgers allow for more efficient and localized consensus models, we argue that they are better suited for the accounting of trade specifications in \ModelName{}.
However, many DAG-based ledgers violate on decentralization.
Since some implementations are not properly tested at scale yet, appending transactions to the ledger is controlled by a centralized coordinator (e.g., IOTA).
Other implementations compromise on decentralization by assembling a fixed group of witness nodes that verify transactions and include them in the ledger (e.g., Nano and Dagcoin).
We specifically aim to avoid authorities with leveraged permissions to prevent (strategic) censorship of the transactions that are submitted to the \ModelName{} blockchain.

\subsection{TrustChain: A Scalable Blockchain for Accounting} \label{sec:trustchain}
Based on the idea of DAG-based blockchains, Otte et al. designed and deployed TrustChain, a blockchain that is optimized for lightweight, tamper-proof accounting of data elements~\cite{otte2017trustchain}.
The key idea is that individuals maintain and grow their \emph{individual ledger} with transactions they have initiated and have been involved in.
Individual ledgers are intertwined (i.e.\ entangled) with each other and form a DAG structure.
TrustChain does not aim to prevent fraud like the double spend attack but instead guarantees eventual \emph{detection} of such fraud.
This yields superior scalability compared to other ledgers but allows for the situation where some fraud targeted at the blockchain layer might go undetected for some time.
Individuals in TrustChain are not required to store all transactions in the network and might choose to store different parts of the global DAG ledger.
Consensus is only reached between interacting parties and not on a global scale.

\begin{figure*}[t]
	\centering
	\begin{subfigure}[t]{.33\textwidth}
		\centering
		\captionsetup{width=.9\linewidth}
		\includegraphics[width=.35\linewidth]{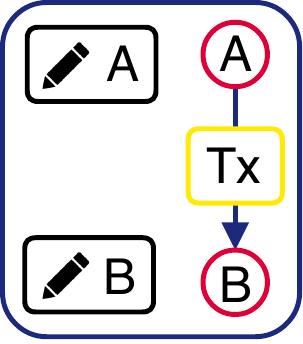}
		\caption{A block with a transaction (\emph{Tx}) between two users ($ A $ and $ B $). Each block contains a single transaction, two digital signatures and two digital identities.}
		\label{fig:trustchain_tutorial_1}
	\end{subfigure}%
	\begin{subfigure}[t]{.33\textwidth}
		\centering
		\captionsetup{width=.9\linewidth}
		\includegraphics[width=\linewidth]{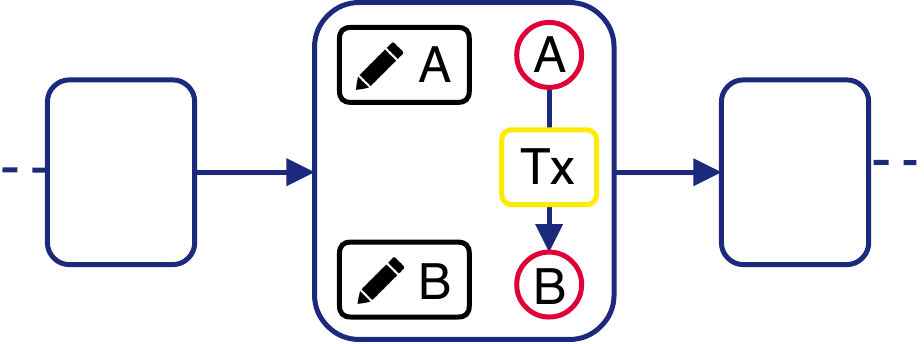}
		\caption{A blockchain of transactions. Each block in the chain contains a hash that describes the prior block.}
		\label{fig:trustchain_tutorial_2}
	\end{subfigure}%
	\begin{subfigure}[t]{.33\textwidth}
		\centering
		\captionsetup{width=.9\linewidth}
		\includegraphics[width=.7\linewidth]{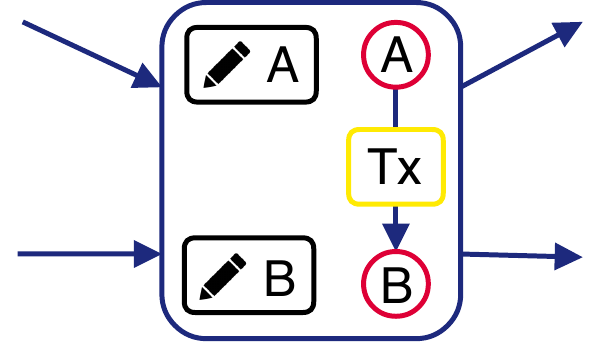}
		\caption{To increase security, each block also references the previous block in the chain of the other transaction participant.}
		\label{fig:trustchain_tutorial_3}
	\end{subfigure}
	\par\bigskip
	\begin{subfigure}[t]{.33\textwidth}
		\centering
		\captionsetup{width=.9\linewidth}
		\includegraphics[width=.7\linewidth]{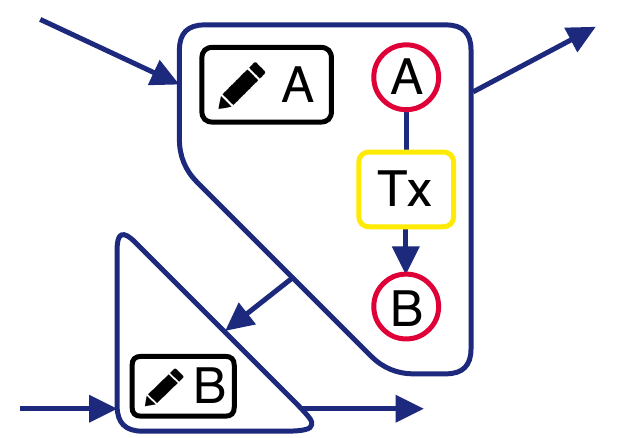}
		\caption{To improve scalability, we extend the TrustChain structure to support concurrent block creation.}
		\label{fig:trustchain_tutorial_4}
	\end{subfigure}\hspace{0.05\textwidth}%
	\begin{subfigure}[t]{.33\textwidth}
		\centering
		\captionsetup{width=\linewidth}
		\includegraphics[width=\linewidth]{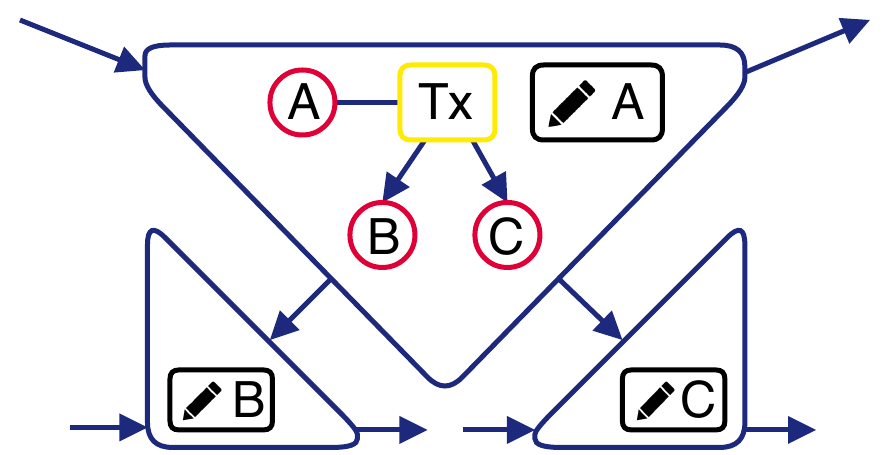}
		\caption{By extending the number of pointers and signatures, we add support for multi-party agreements.}
		\label{fig:trustchain_tutorial_5}
	\end{subfigure}
	\caption{Recording transactions in TrustChain.}
	\label{fig:trustchain_tutorial}
\end{figure*}

We argue that TrustChain is a suitable blockchain to use during the \ModelName{} trading protocol for the following five reasons.
First, it is a blockchain that is optimized for efficient and tamper-proof accounting of generic data items within transactions.
The accounting capabilities of TrustChain enable \ModelName{} to securely persist trade specifications.
Second, TrustChain does not require network-wide replication of all transactions but enables individuals to selectively share parts of their individual ledger with others.
This highly reduces storage requirements and allows \ModelName{} to also run on devices with storage limitations.
Third, TrustChain does not require any trusted authority for the creation, validation, or storage of transactions.
This satisfies our requirement for decentralization, avoiding any reliance on trusted third parties during the \ModelName{} trading protocol.
Forth, the TrustChain structure is optimized for bilateral interactions that are signed by two parties but also supports the creation of unilateral transactions.
This aligns well with the \ModelName{} trading protocol since several operations benefit from support for bilateral transactions (e.g., trade agreements).
Finally, TrustChain is already being used by various decentralized applications that require scalable accounting~\cite{pouwelse2017laws}.
At the time of writing, the global TrustChain ledger contains over 64 million transactions, created by 41.098 unique identities.\footnote{http://explorer.tribler.org}


\subsection{Recording TrustChain Transactions}
We now outline how a transaction between two interacting users $ A $ and $ B $ is recorded in TrustChain, see Figure \ref{fig:trustchain_tutorial}.
Figure \ref{fig:trustchain_tutorial_1} highlights one block containing a transaction between $ A $ and $ B $.
Each block contains a single transaction (\emph{Tx}).
A transaction is a generic description of any interaction between users, for instance, making an agreement or recording a trade.
Both transacting parties digitally sign the transaction they are involved in by using any secure digital signing algorithm.
These signatures are included in the block and ensure that participation by both parties is irrefutable.
It also confirms that both parties agree with the transaction itself.
Digital signatures can be effectively verified by others.
After all required signatures have been added to a record, the transaction is committed to the local databases of the two interacting parties.

The security of stored blocks is improved by linking them together, incrementally ordered by creation time.
In particular, each block is extended with a description (hash) of the prior block.
Each block has a sequence number that indicates its position in the individual ledger.
This yields the blockchain structure as shown in Figure \ref{fig:trustchain_tutorial_2}.
As a result, each user maintains their own local chain which contains all transactions in which they have participated.
This sets TrustChain apart from the structure of traditional blockchains, where the entire network maintains a single, linear ledger.

Note how the blockchain structure in Figure \ref{fig:trustchain_tutorial_2} allows user $ A $ to modify blocks in their individual ledger without being detected by others.
In particular, $ A $ can reorder the blocks in its individual ledger since validity can quickly be restored by recomputing all hashes.
In most blockchain applications, the global consensus mechanism prevents this kind of manipulation.
TrustChain uses a more efficient approach: each block is extended with an additional (hash) pointer that points to the prior block in the individual ledger of the transaction counterparty.
This is visualized in Figure \ref{fig:trustchain_tutorial_3}.
Each block now has exactly two incoming and two outgoing (hash) pointers, except for the last block in an individual ledger, which only has two incoming pointers.
Modifications of the individual ledger by $ A $, like reordering or removing blocks, can now be detected by a transaction counterparty.
To prove this fraud, the transaction counterparty reveals both the correct block and the invalid block created by $ A $.

\begin{figure}[t]
	\centering
	\includegraphics[width=.6\linewidth]{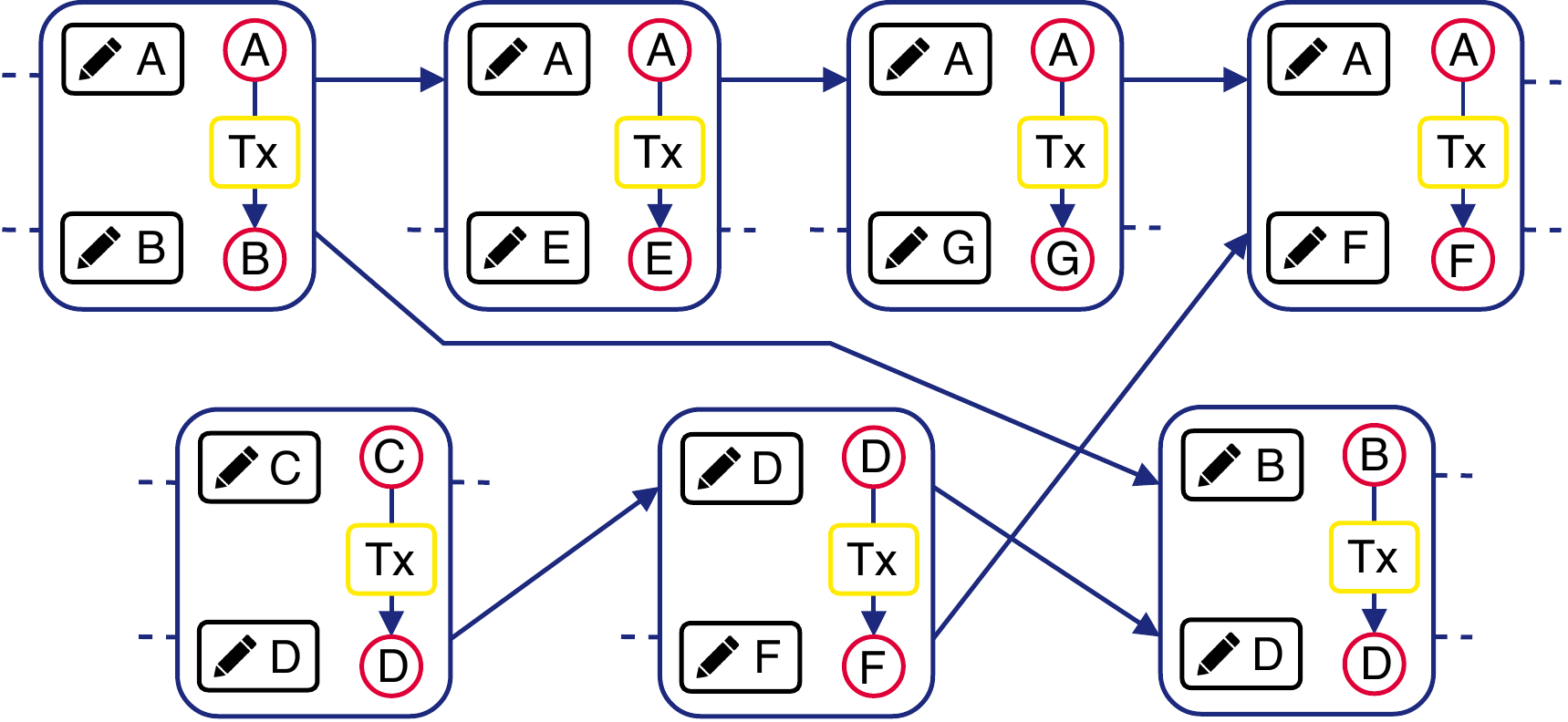}
	\caption{The TrustChain ledger, with seven blocks created by seven participants.}
	\label{fig:trustchain}
\end{figure}

When two parties transact and create a block, their chains essentially become entangled.
When users initiate more transactions with others, it leads to the directed acyclic graph (DAG) structure as shown in Figure \ref{fig:trustchain}.
Figure \ref{fig:trustchain} shows seven blocks, created by seven unique users.
Each block is added to the local chains of all parties involved in the transaction exactly once.
For a more advanced analysis of the technical specifications and security of TrustChain, we refer the reader to the original paper by Otte et al.~\cite{otte2017trustchain}.

\subsection{Improving TrustChain Scalability}
According to Otte et al., TrustChain is designed to scale~\cite{otte2017trustchain}.
However, we identify that its design limits a user to one pending block creation at once.
The main issue is that the digital signature of the transaction counterparty is required before a new block can be appended to an individual ledger (since the input for the hash of each new block includes all signatures in the previous block).
This enables an attack where a malicious user can purposefully slow down the block creation of others by delaying the signing process of a bilateral transaction it is involved in.
It also limits the growth rate of individual ledgers and reduces the overall scalability of TrustChain.

We contribute to TrustChain and improve its scalability by adding support for \emph{concurrent block creation}.
The idea is to remove the requirement for a digital signature of the counterparty when appending new blocks to an individual ledger.
We believe that this concurrency is necessary since it allows traders to engage in transactions with different parties while being engaged in a trade.

Our solution is visualized in Figure \ref{fig:trustchain_tutorial_4}.
It shows a transaction between users $ A $ and $ B $, initiated by $ A $.
We partition a block in two parts, and each block partition is appended to the individual ledger of exactly one party.
Construction of a block between $ A $ and $ B $ now proceeds as follows: first, user $ A $ initiates a transaction by constructing a block partition with the transaction content and its digital signature.
User $ A $ adds this block partition to its individual ledger immediately (note that it does not include the digital signature of $ B $).
$ A $ now sends the block partition to $ B $.
If $ B $ agrees with the transaction, it signs the block partition created by $ A $, adds it to its individual ledger and sends his block partition (with their signature) back to $ A $.
User $ A $ stores the block partition created by $ B $ in its local database.
Participation of both parties in this transaction can now be proven with both block partitions.
This mechanism allows users to be involved in multiple transactions and block constructions at once.

We further improve upon the TrustChain design and enable support for multi-party agreement, see Figure \ref{fig:trustchain_tutorial_5}.
By extending the number of pointers and signatures, we can securely record transactions between more than two parties.
Block partitions can now be represented as a tree, where the root node is the block partition created by the transaction initiator.
While we do not currently use this mechanism in \ModelName{}, we believe this allows for more advance features like support for $n$-way trades~\cite{wang2018loopring}.

\subsection{Storing Trade Specifications on TrustChain}
\begin{figure}[t]
	\centering
	\includegraphics[width=.8\linewidth]{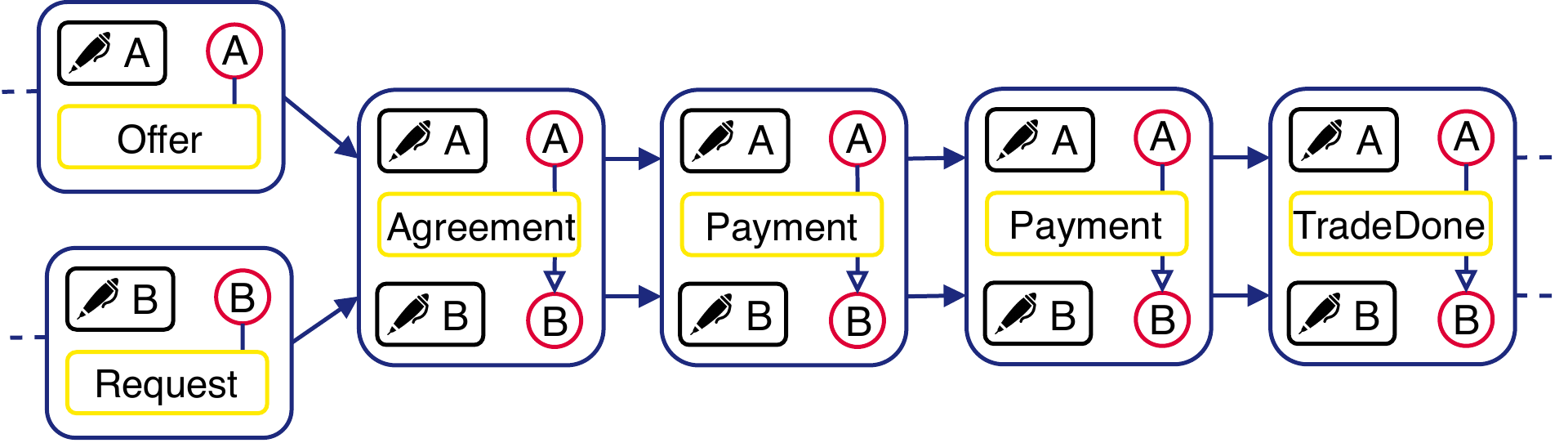}
	\caption{A part of the TrustChain ledger, storing an offer and request created by traders $ A $ and $ B $ respectively, and full specifications of a finished trade between $ A $ and $ B $.}
	\label{fig:trustchain_market}
\end{figure}

We now outline how trade specifications are stored within transactions on TrustChain.
Figure \ref{fig:trustchain_market} shows a part of the TrustChain ledgers of traders $ A $ and $ B $.
It includes a sequence of transactions that indicate a finished trade between $ A $ and $ B $.
Trade agreements, created during phase III in the \ModelName{} trading protocol, are stored within a bilateral \TRAgreement{} transaction which is digitally signed by both involved traders.
Individual payments are stored within bilateral \TRPayment{} transactions.
A \TRPayment{} transaction that indicates an asset transfer from trader $ A $ to $ B $ is only signed by $ A $ if the assets are successfully transferred, and only signed by $ B $ if the assets are observed in one of their wallets.
Finally, trade finalizations are stored within bilateral \TRTradeDone{} transactions.

Since the overhead of creating new TrustChain transactions is low, we also store offers and requests as unilateral transactions.
Figure~\ref{fig:trustchain_market} shows a \TROffer{} transaction, created by user $ A $, and a \TRRequest{} transaction, created by user $ B $.
Each of these transactions only contain a digital signature of its creator.
Note that new orders stored on individual ledger of a specific peer are not locked in time until this peer initiates a bilateral transaction with another peer.

\section{Implementation and Evaluation}
\label{sec:evaluation}
We present the open-source implementation of \ModelName{} and an experimental evaluation.
The evaluation answers the following three questions: (1) what is the duration of a single trade when \ModelName{} is deployed on low-resource devices? (2) How scalable is \ModelName{} in terms of throughput and bandwidth usage when increasing the system load? And (3) how does the scalability of \ModelName{} compare with that of state-of-the-art decentralized asset exchanges?

\subsection{Implementation Details}
\label{sec:implementation}
We have implemented the \ModelName{} trading protocol and system architecture in the Python 3 programming language.
The implementation is open source and all software artifacts (source code, tests, and documentation) are published on GitHub.\footnote{https://github.com/tribler/anydex-core}

\textbf{Networking.}
We have built \ModelName{} on top of an existing networking library
which provides the functionality to devise decentralized overlay networks and has built-in support for authenticated network communication, custom message definitions, and UDP hole punching.\footnote{https://github.com/tribler/py-ipv8}
For efficiency reasons, the UDP protocol is used for message exchange between peers.
The \ModelName{} implementation uses request stores to handle message timeouts and transmission errors.

\textbf{Architecture.}
We have implemented all components of the \ModelName{} system architecture, presented in Figure~\ref{fig:mechanism_architecture}.
Request stores are organized in a dictionary with the request identifiers as keys and the individual request stores as values.
The matching priority queue is implemented as a linked list that remains sorted whenever new entries are added to it.
Different order books are organized in a dictionary whose keys are an asset pair, and the values are the individual order books.
We provide both an implementation of a basic order book and an implementation of a limit order book.
The basic order book structure can be used to store and match orders with any attributes.
The limit order book is optimized to store asset trading orders with pricing information and organizes orders with the same price in a so-called \emph{price level}.
It traverses through the set of orders based on these price levels during order matching.

\textbf{Wallets.}
Our implementation contains a \texttt{Wallet} base class that can be extended by programmers to create wallets that store different types of assets.
For testing purposes, we have implemented a \texttt{DummyWallet}, which is used when executing the unit tests and when running our experiments.
This wallet does not interact with any settlement provider.
For practical reasons, we also provide a \texttt{BitcoinWallet} that invokes the \emph{bitcoinlib} library to query Bitcoin balance or to transfer Bitcoin to other wallets.\footnote{http://github.com/1200wd/bitcoinlib}
This library communicates with full nodes operating in the Bitcoin network.

\textbf{Policies.}
The implementation contains base classes for the different policies that \ModelName{} requires.
Developers can implement their own advanced policies by subclassing these base classes.
For example, programmers can add custom matching policies by subclassing the \texttt{MatchPolicy} class and by providing an implementation of the \texttt{match(order, order\_book)} method.
This method accepts the order being matched and an order book as parameters and returns a list of orders that matches the passed order.
We provide an implementation of the \emph{price-time} matching policy.
This policy is used when matching assets that contain a price expressed in terms of another asset.
The experiments in this section use this matching policy to match new orders.
We implement and use a basic validation policy that checks whether an incoming order has not expired yet.
Finally, we provide a basic order dissemination policy where a new order is sent to the same subset of matchmakers every time.

\subsection{Trading on Low-resource Devices}
\label{sec:exp_trading_low_devices}
The first experiment determines the total duration of a trade between two traders when running \ModelName{} on low-resource devices.

\textbf{Setup.}
This experiment is conducted with two hosted Raspberry Pis (3rd generation, model B+).
The devices run the Raspbian Stretch operating system and the Python 3.5 interpreter.
Each device creates an order, and the two created orders indicate mutual interest in a trade.
One device assumes the identity of trader $ A $ and the other device acts as trader $ B $.
To avoid the transmission of redundant \MsgMatch{} messages, only one of the two devices acts as a matchmaker.
The experiment is executed in an isolated environment: there is only network communication between the two Raspberry Pis.
For this experiment, we use two different subclasses of \emph{DummyWallet}.
We configure these wallet such that assets instantly arrive when being transferred to another wallet.
We do so to measure the overhead of \ModelName{} itself, and not that of asset transfer operations.

During the experiment, we log the timestamp of several events. 
At $ t=0 $, both orders are created.
The trade is finished when both trading parties have signed a \TRTradeDone{} transaction and have committed the transaction to their individual ledgers.

\begin{figure}[t]
	\centering
	\includegraphics[width=\linewidth]{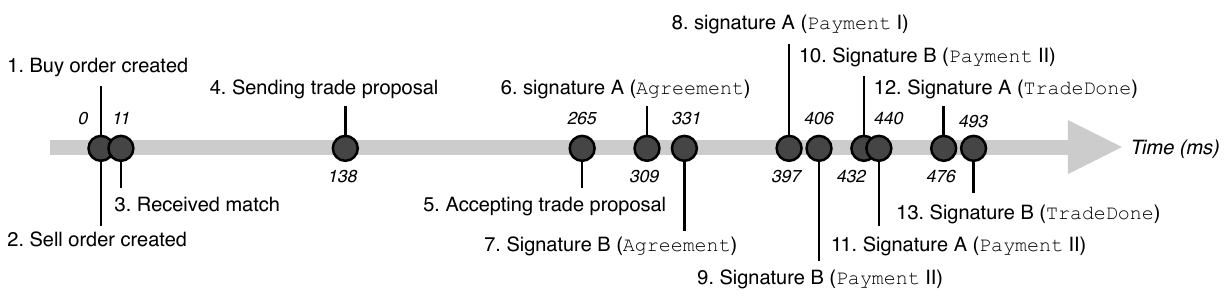}
	\caption{A timeline of the events during a single trade between two traders $ A $ and $ B $. The experiment is conducted on two hosted Raspberry Pis (3rd generation, model B+). The total duration of the trade is 493 milliseconds.}
	\label{fig:trade_timeline}
\end{figure}

\textbf{Results.}
Figure \ref{fig:trade_timeline} shows a timeline of the events during a single trade between the two Raspberry Pis.
The full trade sequence, from the moment of order creation to mutual possession of a dual-signed \TRTradeDone{} transaction, completes in 493 milliseconds, less than half a second.
Almost half of the trade duration, 254 milliseconds, is spent in phase III of the \ModelName{} trading protocol, the trade negotiation phase.
During this phase, a trader determines whether a matched counterparty is already involved in a trade by inspection of the entries in their TrustChain ledger.
Note that at the start of the trade negotiations, both traders have a single transaction in their individual TrustChain ledger, containing specification of their created order.

\textbf{Conclusion.}
This experiment shows that a full trade, including order creation, can be completed within half a second on low-resource devices if asset transfer would be instant.
Since orders are published as unilateral transactions on TrustChain, order creation is almost instant.
This duration is significantly lower than the block creation times of state-of-the-art DEXes.
For example, the average block creation time of the NEM blockchain platform is around 60 seconds on average~\cite{nem2018nem}.
This means that it can take up to a minute before a new order is confirmed on the NEM blockchain.

An implementation of \ModelName{} in a real Internet-of-Things environment would be viable since its communication and transaction creation overhead is minimal compared to existing decentralized asset trading marketplaces (see Section~\ref{subsec:scalability_experiment}).
However, our trading protocol requires periodic queries of other blockchain ledgers during an ongoing trade.
Since maintaining the full transaction history is not realistic given the storage restrictions of IoT devices, \ModelName{} should rely on dedicated full nodes that maintain synchronized with all transactions on a specific blockchain.
We believe that devices with less processing capabilities than the Raspberry Pis used during this experiment are capable of maintaining and securing individual TrustChain ledgers.
This should be verified with a small-scale deployment of \ModelName{} in an IoT environment where blockchain-based assets are managed.

While the low trade duration on low-resource devices is a promising result, the experiment is not representative of a realistic trading environment where there are many traders creating orders and exchanging assets simultaneously.
Therefore, our next experiment focuses on the scalability of \ModelName{} and reveals how our mechanism behaves under a higher system load.

\subsection{Scalability of \ModelName{}}
\label{subsec:scalability_experiment}
We now perform scalability experiments to quantity the performance of \ModelName{} as the system load and network size increases.

\textbf{Setup.}
To explore the limitations of \ModelName{}, we conduct scalability experiments on our university cluster.
The detailed specifications of the hardware and runtime environment can be found online.\footnote{https://www.cs.vu.nl/das5/}
Our infrastructure allows us to reserve computing nodes and deploy instances of \ModelName{} on each node.
We use the Gumby experiment framework to orchestrate the deployment of \ModelName{} instances onto computing nodes and to extract results from experiment artifacts.\footnote{https://github.com/tribler/gumby}
The scalability experiment is controlled by a \emph{scenario file}, a chronologically ordered list of actions which are executed by all or by a subset of running instances, at specific points in time after the experiment starts.
Each run is performed at least five times, and the results are averaged.

We increase the system load, namely the number of new orders being created every second.
As the system load grows, so does the number of traders in the network.
With a system load of $ l $ new orders per second, we deploy $ 0.5l $ \ModelName{} instances.
We devise a synthetic dataset to determine the performance of \ModelName{} under a predictable arrival rate of orders.
In a network with $ n $ peers running \ModelName{}, $ n $ orders are created every half a second.
Each peer alternates between the creation of an offer and a request.
To avoid the situation where all instances create new orders at the same time, the starting time of this periodic order creation is uniformly distributed over all peers, based on their assigned IDs (ranging from 1 to $ n $).
Each peer acts as a matchmaker and sends a new order to four matchmakers, which each peer randomly selects when the experiment starts.
The experiment lasts for 30 seconds, after which $ 60n $ orders are created in total.
We fix the price of each new offer and request to \$1 and the asset quantities to 1, to make matchmaking a predictable process.
After 30 seconds, the experiment is terminated.

We test the scalability of \ModelName{} under a combination of different risk mitigation strategies.
The $ RESTRICT(t) $ policy denotes the policy where a trader will not trade with a counterparty that currently holds responsibility in $ t $ or more ongoing trades (see Section \ref{sec:limit_risk}).
With the $ INC\_SET(n) $ policy, we refer to the incremental settlement strategy where each trader makes $ n $ payments to the counterparty during a single trade.
We consider four experiment settings in total, with combinations of the $ RESTRICT(1) $ and $ INC\_SET(2) $ policies, and when no risk mitigation policy is active.
Scalability is measured as follows:
first, we analyse the peak \emph{throughput} observed during the experiment, in terms of trades per second.
Second, we consider the average order fulfil \emph{latency}, which is the time between the creation of an order and the time until this order has been completed (the order creator has exchanged all assets as specified in the order).

\begin{figure*}[t]
	\centering
	\begin{subfigure}[t]{.5\textwidth}
		\centering
		\captionsetup{width=.9\linewidth}
		\includegraphics[width=.95\linewidth]{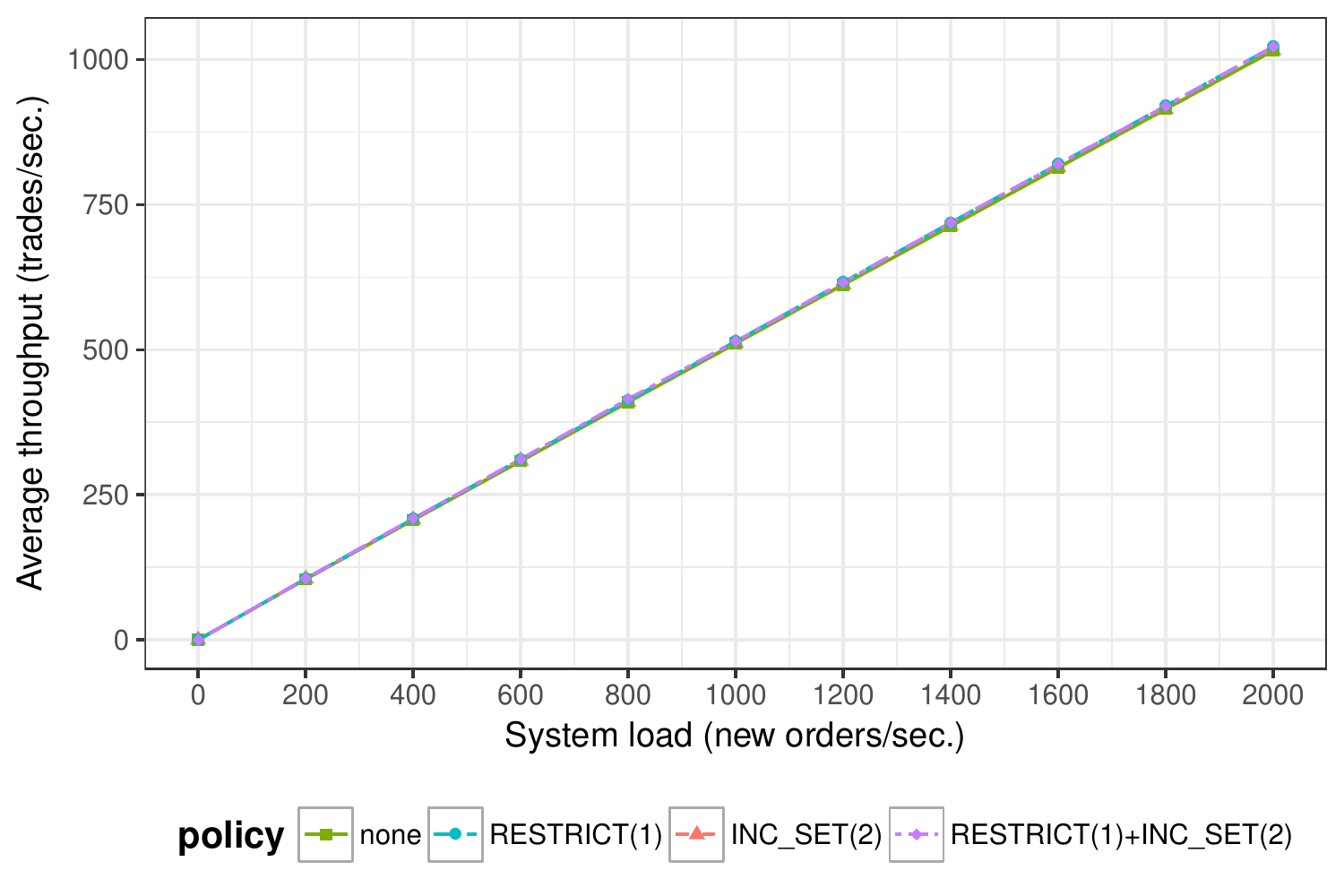}
		\caption{Peak throughput}
		\label{fig:scalability}
	\end{subfigure}%
	\begin{subfigure}[t]{.5\textwidth}
		\centering
		\captionsetup{width=.9\linewidth}
		\includegraphics[width=.95\linewidth]{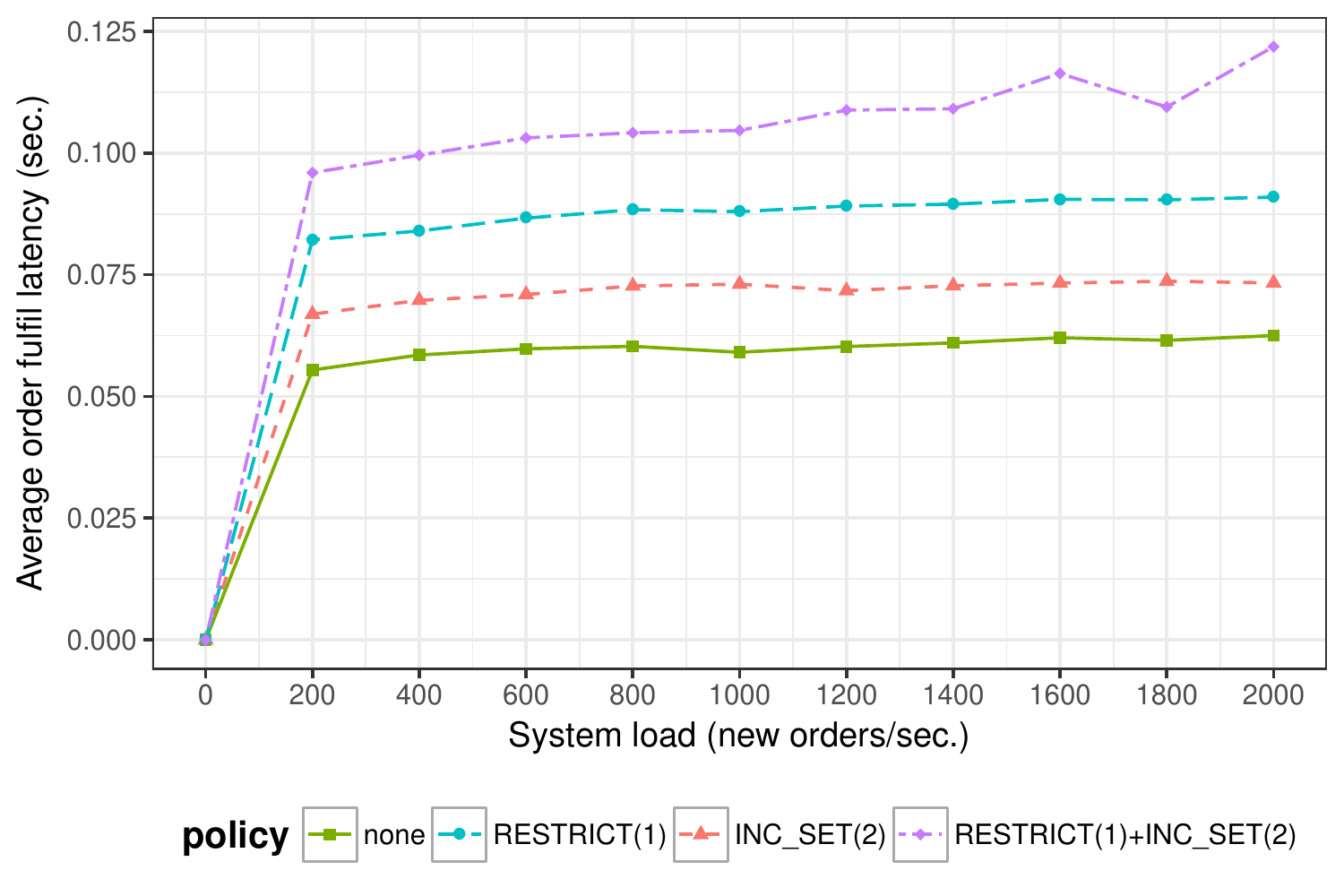}
		\caption{Average order fulfil latency}
		\label{fig:latency}
	\end{subfigure}%
	\caption{The peak throughput and order fulfil latencies as the system load increases.}
	\label{fig:scalability_latency}
\end{figure*}


\textbf{Results.}
The results of the scalability experiments are presented in Figure~\ref{fig:scalability_latency}.
We run each experiment with a specific system load up to 1.000 deployed instances (which is close to the limitations of the used hardware).
Figure~\ref{fig:scalability} shows how the peak throughput (expressed in trades per second, vertical axis) behaves with respect to the system load (horizontal axis).
All experiment settings hint at linear scalability as the system load increases.
Furthermore, enabling risk mitigation policies does not appear to have a notable effect on the peak throughput.
Experimentation on more compute nodes should reveal whether this trend continues when the system load exceeds 2.000 new orders per second.

Figure~\ref{fig:latency} shows the average order fulfil latency when the system load increases, for the four evaluated experiment settings.
The average order fulfil latency remains largely constant when the system load grows.
Applying the restriction and incremental settlement policies increases the average order fulfil latency, since more operations have to be performed to successfully complete an order.
We observe that there is a moderate increase of latency when applying the $ RESTRICT(1) + INC\_SET(2) $ policies when the system load grows to 2.000 trades per second.
The high system load is likely to increase the duration of individual trades beyond 0.5 seconds, which means that the $ RESTRICT(1) $ policy prevents traders from initiating a new trade with others.
Since a trader now has to find a new party to trade with, the average order fulfil latency increases.


\textbf{Conclusion.}
The main finding of this experiment is that the throughput (trades per second) scales linearly with respect to the system load and network size.
We also observe that the average order fulfil latency remains largely constant as the system load grows.
Further experimentation should reveal whether these trends continue with a higher order creation rate.

\subsection{Scalability comparison}
\label{sec:experiment_comparison}
We compare the scalability of \ModelName{} with that of state-of-the-art DEXes.
This work is the first to present a systematic scalability comparison between DEXes, to the best knowledge of the authors.
For the following experiment, we compare \ModelName{} with the BitShares and Waves DEXes respectively.
BitShares and Waves have attained significant market capitalization (\$80 million and \$83 million respectively), and have employed consensus mechanisms that are fundamentally different compared to the TrustChain blockchain used by \ModelName{}.
We briefly discuss BitShares and Waves.

\textbf{BitShares.}
BitShares enables users to issue, manage and trade their own assets, which are stored on a single blockchain~\cite{schuh2015bitshares}.
To coordinate block creation, BitShares uses the Delegated Proof-of-Stake (DPoS) consensus mechanism~\cite{larimer2014delegated}.
DPoS utilizes approval voting to decide on a committee of so-called \emph{witnesses}.
Witnesses are able to append a new block to the blockchain (produce a block), in a round-robin fashion, and are rewarded by the sum of transaction fees in a block they have added to the ledger.
If a witness acts malicious, i.e. by deliberately failing to produce a new block, the witness will eventually be removed from the committee by stakeholders.
All orders are stored on the BitShares blockchain.
Matching of orders proceeds by a deterministic algorithm during transaction validation.
Specifications on resulting matches are not stored on the blockchain, to improve efficiency (but they can be deduced by replaying transactions).

We compile BitShares version 3.3.2.
During our experiments, we deploy a committee with 21 witnesses that produce blocks, comparable to the committee in the BitShares mainnet at the time of writing.
We fix the block creation interval to five seconds.

\textbf{Waves.}
Similar to BitShares, the Waves platform also allow users to create and manage custom assets~\cite{wavesplatform}.
The adopted consensus mechanism is Waves-NG, a protocol that combines Proof-of-Stake and Bitcoin-NG~\cite{eyal2016bitcoin}.
Although the blockchain in Waves is a decentralized data structure, order matchmaking in Waves is largely centralized and traders submit new orders to a single matchmaking instance.
The submitted order can then only be matched with other orders this matchmaker knows about.
When two orders are matched by a matchmaker in Waves, the resulting assets are exchanged on the blockchain and this trade is then recorded as a transaction.
The matchmaker collects the transaction fees in both matched orders.

We compile Waves version 1.1.5 and adopt all default settings when deploying a Waves development network.
To ensure a relatively quick block creation, the average time between creation of two blocks is lowered to five seconds.
We configure each deployed Waves instance to run a matchmaker.
Each Waves instance sends new orders to another Waves instance, which remains fixed throughput the experiment.

\textbf{Setup.}
Similar to the prior experiment, we increase the system load and observe the peak throughput and average order fulfil latency of each system.
For each run with BitShares or Waves, we initiate the experiment by starting a single BitShares/Waves instance, which creates two new types of digital assets on the blockchain.
This instance then transfers sufficient assets to all other instances, so they can create new orders and trade these assets with others.
When each instance has sufficient funds to create numerous orders, they start to create a new order every half a second, for one minute in total.
We adopt the same synthetic workload as in the previous scalability experiment.
For \ModelName{}, we enable the $ RESTRICT(1) $ and $ INC\_SET(2) $ risk mitigation policies.

In the prior experiments, we have adopted the peak number of trader per second as our scalability metric.
In order to fairly compare the scalability between \ModelName{}, BitShares, and Waves, we modify the throughput metric and quantify the peak number of \emph{operations} per second observed during the experiment.
For BitShares, we count the number of operations included in a transaction.
For Waves, we consider a new transaction on the blockchain as a single operation. 
For XChange, we consider the creation of a block in an individual ledger as a single operation. 

\begin{figure*}[t]
	\centering
	\begin{subfigure}[t]{.5\textwidth}
		\centering
		\captionsetup{width=.9\linewidth}
		\includegraphics[width=.95\linewidth]{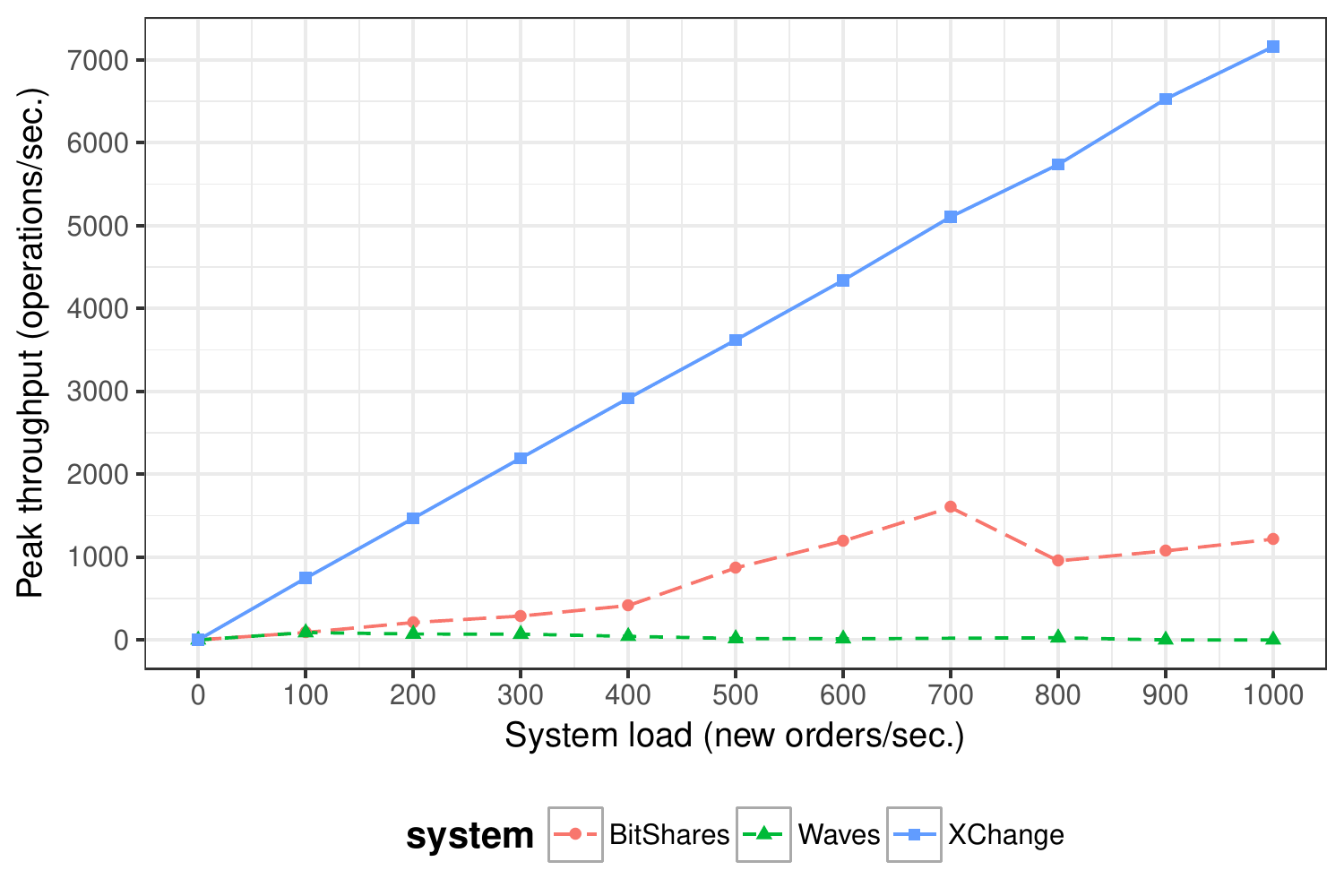}
		\caption{Peak throughput}
		\label{fig:scalability_comparison}
	\end{subfigure}%
	\begin{subfigure}[t]{.5\textwidth}
		\centering
		\captionsetup{width=.9\linewidth}
		\includegraphics[width=.95\linewidth]{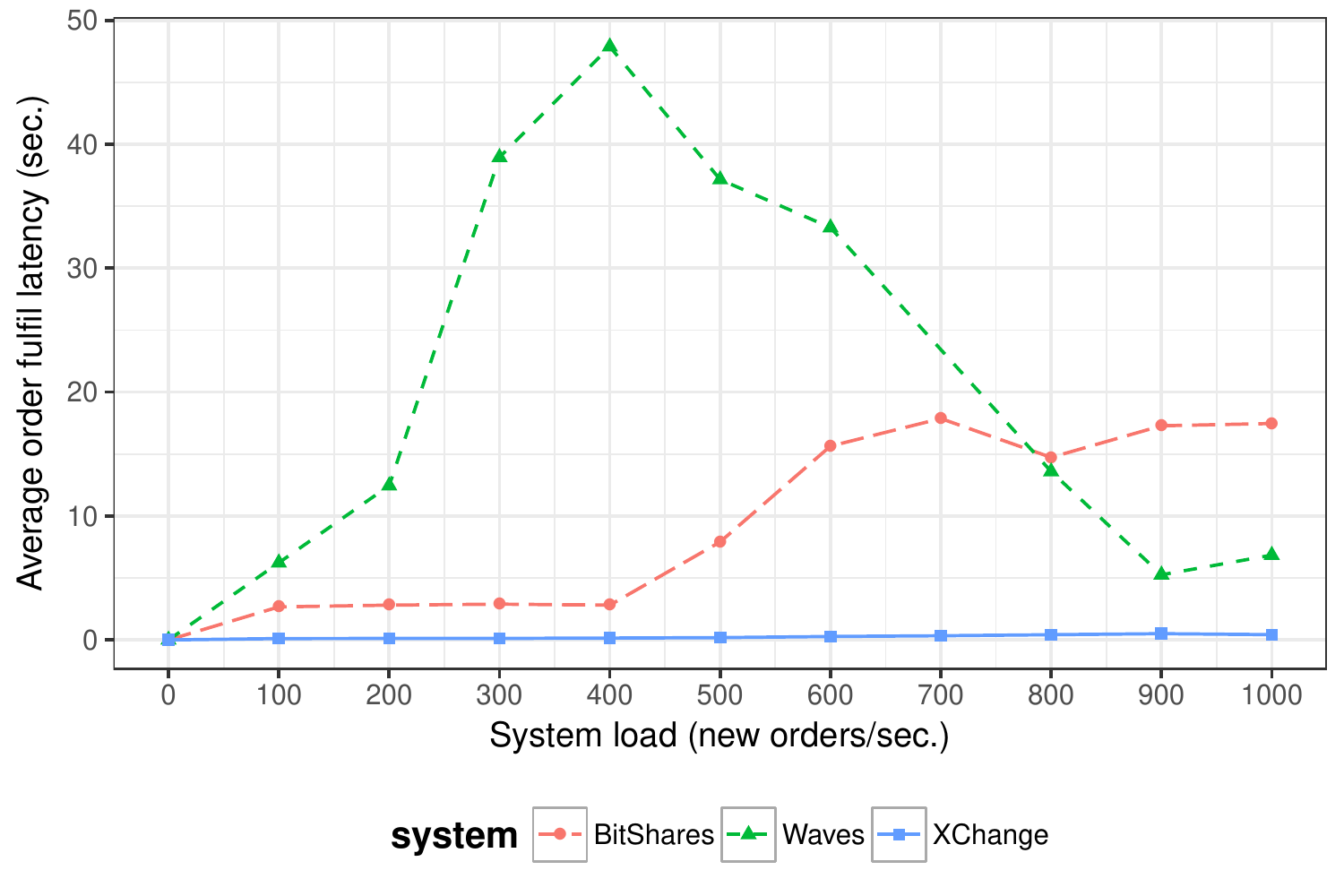}
		\caption{Average order fulfil latency}
		\label{fig:latency_comparison}
	\end{subfigure}%
	\caption{The peak throughput and order fulfil latencies of \ModelName{}, BitShares and Waves, as the system load increases.}
	\label{fig:scalability_latency_comparison}
\end{figure*}

\textbf{Results.}
Figure~\ref{fig:scalability_latency_comparison} shows the scalability of \ModelName{}, BitShares and Waves.
Figure~\ref{fig:scalability_comparison} shows how the peak throughput in terms of operations per second behaves as the system load increases.
The peak throughput of Waves does not exceed 89 operations per second.
Furthermore, we observe that a majority of the deployed Waves instances crashes due to memory limitations when increasing the system load to 300 new orders per second.
The consensus mechanism of Waves is not able to keep up with the increasing system load.
Up to a system load of 400 new orders per second, BitShares shows a linear increase in peak throughput.
Surprisingly, when the system load exceeds 400 new orders per second, peak throughput grows faster.
We found that BitShares has issues keeping up with a higher load, and the inclusion of orders in the BitShares blockchain might be deferred.
Even though the system load is predictable throughout the experiment, the number of transactions inside new blocks on the BitShares blockchain becomes more uneven when the system load grows.
Figure~\ref{fig:scalability_comparison} reveals that \ModelName{} exhibits a near-linear increase in peak throughput.
The TrustChain ledger as used by \ModelName{} achieves the creation of 7.000 new blocks (operations) per second.

Figure~\ref{fig:latency_comparison} shows the average order fulfil latency of \ModelName{}, BitShares and Waves as the system load increases.
The order fulfil latency of Waves increases up to almost 48 seconds, and then starts to decline.
Up to a system load of 400 new orders per second, BitShares shows an average order fulfil latency of around 2.5 seconds.
This is expected since the block creation interval is fixed to five seconds and on average, it takes 2.5 seconds for a new order to be fulfilled.
When the system load grows, so does the average order latency, confirming our finding that the inclusion of incoming orders in a block becomes a less predictable process.
Finally, note how the average order fulfil latency of \ModelName{} remains sub-second during this experiment.

\textbf{Conclusion.}
\ModelName{}, in combination with the TrustChain ledger, shows excellent scalability compared to the BitShares and Waves DEXes.
At this point, we should note that TrustChain, unlike BitShares and Waves, does not incorporate global consensus and therefore has different security guarantees.
Also, during this experiment we assume that \ModelName{} assets are exchanged instantly between trading parties, which is done to quantify the limitations of \ModelName{}.
In reality, a trader most likely has to wait one or more block creation intervals in order to guarantee that a payment is finalized.
This further adds to the order fulfil latency.

\section{Related Work}
\label{sec:related_work}
In this section we report on literature related to the problem domain of \ModelName{}. 
For an extensive discussion on the usage of blockchain-based systems in IoT, we refer the readers to \cite{christidis2016blockchains} and \cite{reyna2018blockchain}. 
We first report on two features related to \ModelName{}, namely 1) auctioning, and 2) decentralizing blockchain-based marketplaces.
We then inform of studies that implement decentralized accountability without network-wide (global) consensus.


\textbf{Auctioning}.
The act of matching traders in \ModelName{} is closely related to the auctioning problem, where buyers and sellers make bids for a specific asset and eventually determine the price of that asset. 
Considerable effort has been spent on the decentralization of the auctioning process, which is the act of conducting an auction without trusted third parties~\cite{fontoura2005decentralized, despotovic2004towards, ogston2002peer, hausheer2005peermart}. 
In \cite{fontoura2005decentralized}, the authors use so-called \textit{law governed interaction} to define rules for auctioning. 
Their framework assumes a single asset type and regulates only one side of the trade, namely sellers. 
In~\cite{despotovic2004towards}, traders regularly broadcast their (buy or sell) prices to all peers in the network and wait for answers from all interested peers.
This is an inefficient way of communication since it floods the network with messages. 
Advanced solutions use so called \textit{brokers} to pre-aggregate market information.
Brokers are similar to matchmakers in the \ModelName{} protocol.
In \cite{ogston2002peer}, peers build clusters and select one leader to be the broker for a specific cluster.
Selecting the cluster sizes raises a trade-off between effectiveness of trade and scalability. 
In PeerMart, a set of brokers is selected deterministically for each service to be traded~\cite{hausheer2005peermart}. 
However, full synchronization among brokers causes significant communication overhead in PeerMart.
We note that our \ModelName{} trading protocol considers asynchronous operation of matchmakers, without the need for round synchronization. 
Unlike all decentralized auctioning protocols mentioned above, \ModelName{} also involves the execution of a trade (the actual asset exchange).

\textbf{Blockchain-based marketplaces}.
\ModelName{} assumes the availability of a blockchain to record trade specifications. 
There have been substantial amount of attempts to design marketplaces that leverage the features of distributed ledgers. 
Beaver combines the features of a public ledger-based consensus protocol and an anonymous payment system for building a privacy-preserving e-commerce system~\cite{soska2016beaver}. 
Similar to \ModelName{}, all marketing activity, as well as payments, are stored on a distributed ledger. 
However, Beaver relies on network-wide consensus on the ledger, while our protocol supports any distributed ledger that has accounting capabilities. 
In~\cite{klems2017trustless}, authors present Desema, a prototype for trading software services. 
Instead of service metadata, it stores references to trade data in the ledger. 
Desema assumes bipartite relationship between users, as each user can either be a service provider or consumer. 
FairSwap considers the fair exchange of digital goods~\cite{dziembowski2018fairswap}. 
It uses smart contracts which are stored in and managed by distributed ledgers.
FairSwap has strong synchronization requirement where every peer in the network is aware of the actual round.
All the works mentioned in this section up to this point consider the trade of goods or services with using a specific asset type (such as Bitcoin) as the medium of exchange. 
\ModelName{}, however, is designed for the trade of \emph{any} asset types.




\textbf{Accountability-oriented distributed ledgers}.
\ModelName{} supports distributed ledgers which do not rely on network-wide (global) consensus for providing accountability.
PeerReview maintains local records of messages and guarantees that deviation of a peer from expected behaviour is eventually detected~\cite{haeberlen2007peerreview}. 
TrustChain uses a similar approach to design a blockchain data structure and provides tamper-proofness by the entanglement of local transaction records~\cite{otte2017trustchain}.
Ongoing research justifies the integrity of these approaches by proving that one does not need consensus to implement a decentralized accounting mechanism~\cite{Guerraoui2019AT2}.

\section{Conclusions}
We have presented \ModelName{}, a blockchain-based mechanism for generic asset trading in resource-constrained environments.
The key idea is to account full trade specifications on a blockchain.
\ModelName{} is generic since it can facilitate cross-chain trade without relying on special transaction types.
Moreover, \ModelName{} is decentralized since it does not rely on trusted third parties to mediate in the trading process, and to bring traders together.
Finally, by careful inspection of the ongoing trades involving a potential trade counterparty, we have limited the effectiveness of counterparty fraud conducted by adversaries.
Incremental settlement further reduces counterparty risk by splitting each payment into multiple smaller ones.

With an open-source implementation and an experimental evaluation, we have demonstrated the viability of trading on devices with low hardware capabilities.
A single trade can be completed within half a second if asset transfers on external blockchain platforms would complete instantly.
With a scalability experiment at scale, we achieved over 1.000 trades per second and found that the throughput of \ModelName{} in terms of trades per second scales linearly with the system load and network size.
Finally, we have compared the scalability of \ModelName{} with that of the state-of-the-art DEXes BitShares and Waves, and conclude that \ModelName{} exhibits superior scalability.



\section*{Acknowledgements}
This work was funded by NWO/TKI grant 439.16.614.


\bibliographystyle{elsarticle-num} 
\bibliography{references}

\end{document}